\documentclass[]{aa}  

\usepackage{natbib}
\bibpunct{(}{)}{;}{a}{}{,}
\usepackage{graphicx}
\usepackage{revsymb}	
\usepackage{amsmath}	
\usepackage[usenames]{color}	
\usepackage{epstopdf}	
\usepackage{txfonts}
\usepackage{amssymb}
\usepackage{color}
\usepackage[justification=centering]{caption}
\usepackage{geometry} 
\usepackage[parfill]{parskip} 
\usepackage{amssymb}
\usepackage{slantsc}
\usepackage{array}
\usepackage{url}
\usepackage{textcomp,    
            xspace}
\usepackage{amsfonts}
\usepackage[colorlinks=true,linkcolor=blue, urlcolor=blue, citecolor=blue]{hyperref}
\usepackage{sidecap}
\usepackage{graphicx}
\usepackage{xcolor}
\usepackage{nicefrac}
\usepackage{subfigure}
\usepackage{epsfig}
\colorlet{rouge}{red!70!darkgray}

\begin{document}
\title{Helioseismic determination of the solar metal mass fraction}
\author{G. Buldgen\inst{1,2} \and A. Noels\inst{2} \and V. A. Baturin\inst{3} \and A. V. Oreshina\inst{3} \and S.V. Ayukov\inst{3} \and R. Scuflaire\inst{2} \and A. M. Amarsi\inst{4} \and N. Grevesse \inst{2,5}}
\institute{D\'epartement d'Astronomie, Universit\'e de Gen\`eve, Chemin Pegasi 51, CH-1290 Versoix, Switzerland \and STAR Institute, Université de Liège, Liège, Belgium \and Sternberg Astronomical Institute, Lomonosov Moscow State University, 119234,Moscow, Russia \and Theoretical Astrophysics, Department of Physics and Astronomy, Uppsala University, Box 516, 751 20 Uppsala, Sweden \and Centre Spatial de Liège, Université de Liège, Angleur-Liège, Belgium}
\date{May, 2023}
\abstract{The metal mass fraction of the Sun $Z$ is a key constraint in solar modelling, but its value is still under debate. The standard solar chemical composition of the late 2000s  have the ratio of metals to hydrogen $Z/X=0.0181$, with a small increase to $0.0187$ in $2021$, as inferred from 3D non-LTE spectroscopy. However, more recent work on a horizontally and temporally averaged  $\textlangle \rm{3D} \textrangle$ model claim $Z/X=0.0225$, consistent with the high values of twenty-five years ago based on 1D LTE spectroscopy.} 
{We aim to determine a precise and robust value of the solar metal mass fraction from helioseismic inversions, thus providing independent constraints from spectroscopic methods.}
{We devise a detailed seismic reconstruction technique of the solar envelope, combining multiple inversions and equations of state to accurately and precisely determine the metal mass fraction value.}
{We show that a low value of the solar metal mass fraction corresponding to $Z/X=0.0187$ is favoured by helioseismic constraints and that a higher metal mass fraction corresponding to $Z/X=0.0225$ are strongly rejected by helioseismic data.}
{We conclude that direct measurement of the metal mass fraction in the solar envelope favours a low metallicity, in line with the 
3D non-LTE spectroscopic determination of 2021. A high metal mass fraction as  measured using a $\textlangle \rm{3D} \textrangle$ model in 2022 is disfavoured by helioseismology for all modern equations of state used to model the solar convective envelope.}
\maketitle
\section{Introduction}

The precise value of the solar metallicity 
has been an issue in solar modelling since the
reappraisal of the carbon, nitrogen and oxygen abundances in the 
2000s
\citep{Allende2001,Allende2002,Asplund2004,Asplund2006,Melendez2008}. The downwards revision of these and other elements reflects an improved understanding of the solar spectrum, thanks to the development of three-dimensional (3D) radiative-hydrodynamic simulations of the solar photosphere, as well as more complex model atoms for taking account departures from local thermodynamic equilibrium (LTE). These abundance revisions were summarised in \citet[][hereafter AGSS09]{AGSS09}. Further improvements were made over the years  \citep{Scott2015,Scott2015II,Grevesse2015,Amarsi2017,Amarsi2018,Amarsi2021}
culminating in a new solar abundance table presented in \citet[][hereafter AAG21]{Asplund2021} resulting from the best 3D non LTE analyses available for a very large number of elements. The authors find a metal mass fraction of $Z=0.0139$, or $Z/X=0.0187$ relative to hydrogen.

These abundance revisions have led to strong disagreements with classical helioseismic constraints such as the position of the base of the convective envelope, the helium abundance in the
convective envelope and sound speed inversions, but also with neutrino fluxes \citep[see][ and refs therein]{Bahcall2005,Basu08,SerenelliComp,Buldgen2019R,JCD2021}.

In an attempt to resolve these disagreements, several other spectroscopic analyses of the solar abundances have been presented in the literature over the years. In particular, a series of papers summarised in \citet{Caffau2011} determined a higher metallicity value than in AAG21, corresponding to $Z/X=0.0209$.
Many of these differences were later attributed to the measurement of equivalent widths \citep[e.g.][]{Amarsi2019,Amarsi2020}, 
rather than differences in the 3D models. More details, also of other works, can be found in AAG21.

More recently, \citet{Magg2022} carried out an analysis of the solar flux spectrum using a horizontally and temporally averaged 3D model (herafter $\textlangle \rm{3D} \textrangle$). They determined a high solar metallicity corresponding to $Z/X=0.0225$. This is consistent with the 1D LTE value of
\citet{GS1998}, $Z/X=0.0231$. The authors reported to have solved the so-called ``solar abundance problem'' by bringing back a high metallicity value. 

There are several reasons to be sceptical of this claim.
On the spectroscopic side the analysis of \citet{Magg2022} is based on a $\textlangle \rm{3D} \textrangle$ model.  The authors did not validate their model with respect  to any solar constraints; previous attempts with other $\textlangle \rm{3D} \textrangle$ models illustrate they are vastly inferior to full 3D models \citep{Uitenbroek2011,Pereira2013}. Moreover, their results for their 18 neutral iron lines show a large range and standard deviation of $0.62$ dex and $0.13$ dex (compared to $0.10$ dex and $0.03$ dex respectively in AAG21), suggesting serious deficiencies in their analysis. It should also be noted that their oxygen abundance of $8.77$ dex, based on the blended $630$nm line and the $777$nm triplet that shows strong departures from LTE \citep{Amarsi2016,Amarsi2018} is in disagreement with the values inferred from molecular OH lines \citep{Amarsi2021}.  A more recent determination  of the solar oxygen abundance from the same group \citep{Pietrow2023}, using the center to limb spectra of one OI line, puts the value at $8.73$ dex, in closer agreement to AAG21 as well as other previous studies \citep{Pereira2009, Amarsi2018}.

On the solar modelling side, although the agreement with classical helioseismic constraints is improved with a high metallicity value, this is only the case when using  classical standard solar models with outdated descriptions of the radiative opacities and macroscopic transport \citep[see][for a discussion]{Buldgen2023}. Numerous
studies \citep[see][and references therein for a discussion]{JCD2021} have pointed out that a degeneracy exists between abundances and opacities. As such, the agreement with most helioseismic constraints can just as well be restored by modifications to radiative opacities. Furthermore, taking into account macroscopic transport of chemicals due to the combined effects of rotation and magnetic instabilities could restore the agreement with the seismic helium value in the convective zone
\citep{Eggenberger2022}. Indeed, none of the classical helioseismic constraints provide a direct measurement of the solar metallicity, but rather indirect hints that solar calibrated models using a high metal mass fraction provide a better agreement \citep[see e.g.][]{Buldgen2023}.

To resolve the debate, it may be necessary to measure the solar metallicity in a way that is independent from spectroscopy. This is linked to investigations of the properties of the solar convective envelope and has been attempted as early as 2000 \citep{Baturin2000,Takata2001}, by using the properties of the first adiabatic exponent, $\Gamma_{1}=\frac{\partial \ln P}{\partial \ln \rho}\bigg|_{\rm{S}}$ or the adimensional sound speed gradient, $\frac{r^{2}}{GM_{\odot}}\frac{dc}{dr}$. Further studies were attempted in the early 2000 \citep{Lin2005,Antia2006,Lin2007}, with varying results. These measurements are sensitive to the equation of state of the solar plasma, and the most recent studies using modern equations of state have favoured a low metallicity value \citep{Vorontsov13,VorontsovSolarEnv2014, BuldgenZ}. However, low precisions have been achieved so far, with inferred intervals ranging between $\left[ 0.008, 0.013\right]$. An independent approach based on seismic calibration of solar standard models has inferred a value of $Z=0.0142\pm0.0005$ \citep{Houdek2011}, in good agreement with \citep{Asplund2021}.

In this study, we improve, both in accuracy and precision, the methods used in \citet{BuldgenZ} to infer the chemical composition of the envelope. Our method is based on a combination of seismic reconstruction techniques from \cite{Buldgen2020} and recomputation of the thermodynamic conditions in the solar envelope using both FreeEOS and SAHA-S equations of state. From this detailed analysis, combined with linear inversions of the first adiabatic exponent, we are able to infer precisely the favoured metallicity value in the solar convective envelope. Our results are robust with respect to modifications of the equation of state (EOS) and also indicate that the SAHA-S EOS is favoured over FreeEOS.

\section{Solar models}\label{Sec:Models}

\subsection{Evolutionary and seismic models}\label{Sec:Model}
We start from a set of reference models computed with the Liège Stellar Evolution code \citep{ScuflaireCles} using physical ingredients listed in Table \ref{tabModels1}. The main properties of the models are summarized in Table \ref{tabModels}. We also add the properties of Model MB22-Phot, which is the reference SSM provided in \citet{Magg2022}. Its envelope chemical composition, being the one advised, will be used for comparisons with the inversion results in Sects. \ref{Sec:RealData}. These evolutionary models are computed using an extended calibration procedure, similarly to one model of \citet{Buldgen2023}. All models in this work include macroscopic transport at the BCZ to reproduce the lithium depletion observed at the age of the Sun \citep{Wang2021}.

\begin{table*}[h]
\caption{Physical ingredients of the solar models}
\label{tabModels1}
  \centering
\begin{tabular}{r | c | c | c }
\hline \hline
\textbf{Name}&\textbf{Abundances}&\textbf{EOS}&\textbf{Opacity tables} \\ \hline
Model M1 & MB22 & FreeEOS  & OP \\
Model M2 & MB22 & SAHA-S v7 & OP\\ 
Model A1 & AAG21 & FreeEOS & OPAL\\
Model A2 & AAG21 & SAHA-S v7 & OPAL\\
Model A3 & AAG21 & SAHA-S v3 & OPAL\\
\hline
\end{tabular}
\end{table*}

The extended calibration procedure uses 4 free parameters and 4 constraints, namely the mixing-length parameter, $\alpha_{\rm{MLT}}$, the initial chemical composition (described using the initial hydrogen, $X_{0}$ and metal mass fraction, $Z_{0}$) and an adiabatic overshooting parameter, $\alpha_{\rm{Ov}}$, as free parameters and the solar radius, surface metallicity, solar luminosity and the helioseismic position of the base of the convective envelope $\left(r/R\right)_{\rm{BCZ}}$ as constraints. 

This procedure allows to have an excellent agreement not only in the radial position of the base of the convective envelope, but also on the mass coordinate at the BCZ and the $m_{75}$ constraint (namely the mass coordinate at $0.75$ $\rm{R}_{\odot}$) that \citet{Vorontsov13} described as a key constraint to calibrate the value of entropy in the solar convective zone. 

\begin{table*}[h]
\caption{Envelope properties of the evolutionary solar models}
\label{tabModels}
  \centering
\begin{tabular}{r | c | c | c | c | c | c | c }
\hline \hline
\textbf{Name}&\textbf{$\left(r/R\right)_{\rm{BCZ}}$}&\textbf{$\left( m/M \right)_{\rm{CZ}}$}&m75&\textbf{$\mathit{X}_{\rm{CZ}}$}&\textbf{$\mathit{Y}_{\rm{CZ}}$}&\textbf{$\mathit{Z}_{\rm{CZ}}$}&\textbf{$\mathrm{A}\left( Li\right)$ $\left( dex \right)$} \\ \hline
Model M1&$0.7133$&$0.9759$& $0.9827$ & $0.7319$ & $0.2516$ & $0.0165$& $0.915$\\
Model M2&$0.7133$&$0.9762$& $0.9829$ & $0.7306$ & $0.2530$ & $0.0164$& $0.897$\\ 
Model A1&$0.7133$&$0.9766$& $0.9832$ & $0.7399$ & $0.2463$ & $0.0138$& $0.904$\\
Model A2&$0.7133$&$0.9768$& $0.9834$ & $0.7387$ & $0.2476$ & $0.0137$& $0.904$\\
Model A3&$0.7133$&$0.9769$& $0.9834$ & $0.7386$ & $0.2477$ & $0.0137$& $0.912$\\
\hline
MB22-Phot&$0.7123$&/& / & $0.7394$ & $0.2439$ & $0.0166$& /\\
\end{tabular}
\end{table*}

The models computed from these evolutionary sequences serve as starting points for the construction of seismic models following \citet{Buldgen2020}. These models show a much better agreement in the solar radiative envelope, intrinsically limiting the uncertainties coming from cross-term contributions in the last step of the envelope reconstruction procedure. Moreover, \citet{Buldgen2020} showed that their reconstruction approach significantly improved the agreement in entropy proxy plateau and the density profile in the convective envelope with respect to helioseismic constraints, as well as the $m_{75}$ parameter being after reconstruction within $0.9822\pm0.0002$ from \citet{Vorontsov13} (which is essentially the density profile from a physical point of view and shows that our models would be of high quality also according to their criteria). Therefore, these models provide a robust starting point for detailed investigations of the properties of the solar envelope. The total set also provides adequate conditions for robustness tests on artificial data. More precisely, we used various equations of state to carry out the analysis, as this will be the main source of potential inaccuracies in the models. As we focus only on the solar envelope, the inversion is therefore unaffected by the choice of opacity tables used in the reference models. By construction, the calibrated evolutionary models will have a different chemical composition in the convective envelope as a result of the physical ingredients (e.g. opacities or solar abundances, see e.g. \citet{Buldgen2019} for an illustration). However, through the seismic reconstrution procedure, this effect is erased (as shown in the right panel of Fig. \ref{Fig:G1S}) as the thermodynamical coordinates are then uniquely defined from the inversion and used as inputs for the scan in chemical composition. 

As mentioned above, our main goal is to determine helioseismic constraints on the solar metal mass fraction and compare them to spectroscopic abundance tables. Therefore, we will compare our results to existing tables using acronyms defined as follows: GN93 is \citet{GrevNoels}, GS98 is \citet{GS1998}, AGSS09 is \citet{AGSS09}, AAG21 is \citet{Asplund2021}, MB22 is \citet{Magg2022}. In Table \ref{tabModels}, Models M1 and M2 are built using the recent abundances by \citet{Magg2022}, whereas Models A1, A2 and A3 are built using the \citet{Asplund2021} abundances.

We use two solar frequency datasets to assess the impact of the data on our results. The primary dataset (Dataset 1), used in Sect. \ref{Sec:RealData} and \ref{Sec:EOS} is the one used in \citet{Buldgen2020}, which is a combination of the ``optimal'' dataset of \citet{BasuSun} combined with updated BiSON frequencies of \citet{Davies}. The secondary dataset (Dataset 2) is a combination of BiSON data from \citet{Davies} and \citet{BasuSun} at low degree, but all modes with $\ell > 3$ are taken a 360 days asymmetric fitting from MDI data of \citet{Larson2015}. Inversions are computed using the SOLA inversion technique \citep{Pijpers}, following prescriptions of \citet{RabelloParam} for trade-off parameter calibrations, using the InversionKit software in an adapted configuration. 

\subsection{Envelope properties}\label{Sec:EnvProp}

\begin{figure*}
	\centering
		\includegraphics[width=17cm]{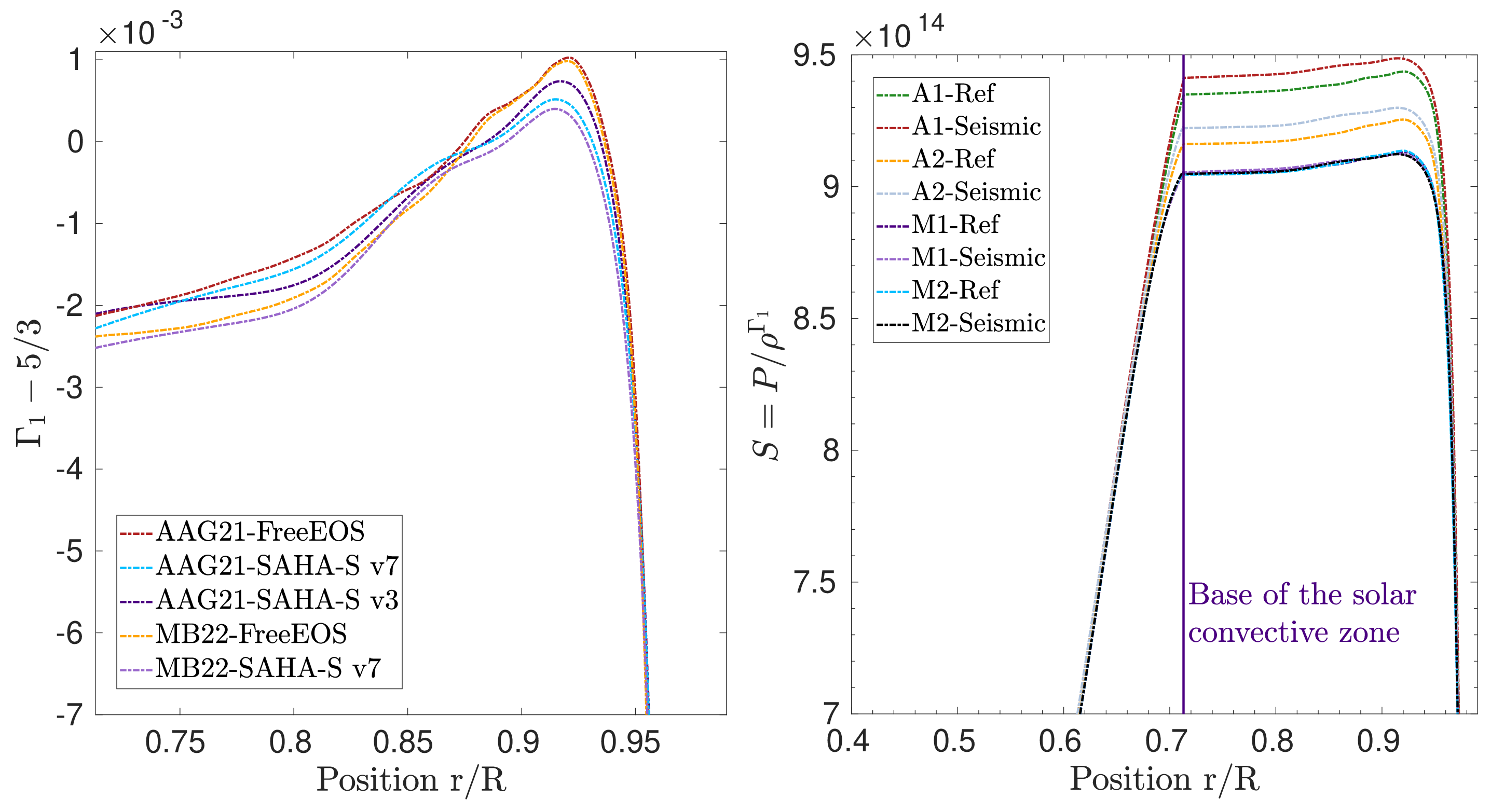}
	\caption{Left panel: $\Gamma_{1}$ profile between the BCZ and $0.95$R$_{\odot}$ for the solar models in Table \ref{tabModels1}. Right panel: Entropy proxy profile, $P/\rho^{\Gamma_{1}}$, as a function of normalized radius for the solar models of Table \ref{tabModels1}. The reference evolutionary models ("Ref") show a significant spread in entropy proxy plateau in the CZ that is corrected in the seismic models ("Seismic") by the seismic reconstruction procedure.}
		\label{Fig:G1S}
\end{figure*}

First, we need to investigate the details of the $\Gamma_{1}$ profile, the entropy proxy profile and the chemical composition in the envelope of the models. We start by plotting in the left panel of Fig. \ref{Fig:G1S} the $\Gamma_{1}$ profile of the evolutionary models and their respective entropy proxy profiles. 

As shown in Fig. \ref{Fig:G1S}, the differences in the lower parts of the convective envelope are minute, and strongly influenced by the EOS. However, it seems that a small trend still exists with metallicity. Namely, the higher the metallicity in the envelope, the lower the $\Gamma_{1}$ value, as a result of the higher signature of the ionization of heavy elements in the case of a higher metal mass fraction. While of the order of $10^{-4}$, these differences may remain significant if the solar data is of high enough quality. These differences can be understood from the point of view of linear perturbations, as it can be shown that relative differences in $\Gamma_{1}$ may be written, assuming a given equation of state, as 
\begin{align}
\frac{\delta \Gamma_{1}}{\Gamma_{1}}=&\left(\frac{\partial \ln \Gamma_{1}}{\partial \ln P}\right)_{\rho,Y,Z}\frac{\delta P}{P} + \left(\frac{\partial \ln \Gamma_{1}}{\partial \ln \rho}\right)_{P,Y,Z}\frac{\delta \rho}{\rho} + \left(\frac{\partial \ln \Gamma_{1}}{\partial Y}\right)_{P,\rho,Z} \delta Y \nonumber \\ 
&+ \left(\frac{\partial \ln \Gamma_{1}}{\partial Z}\right)_{P,\rho,Y} \delta Z, \label{eq:Gamma1Eos}
\end{align}
where we have separated the relative differences in $\Gamma_{1}$ in partial derivatives with respect to the various thermodynamical coordinates of the $\Gamma_{1}$ function. As shown in \citet{BuldgenZ}, the term linked with the Z derivatives has a broad maximum in around $0.75\rm{R}_{\odot}$, which is consistent with the analysis of \citet{Vorontsov13,VorontsovSolarEnv2014}. In practice the inversion procedure will rely on these signatures to determine the solar metallicity from a helioseismic point of view. As already mentioned in previous studies and as can be seen from Fig \ref{Fig:G1S}, the inversion will suffer from a dependency in the EOS. Therefore the metal mass fraction inversion, just like the helium mass fraction inversion, will be affected by the choice of the reference EOS.

In addition to EOS dependencies, it is worth noting that the other terms in Eq. \ref{eq:Gamma1Eos} will vary dependending on the solar model considered. For example, the density and entropy proxy profiles in the envelope of solar evolutionary models strongly vary with the physical ingredients used in the calibration procedure. This is illustrated in Fig \ref{Fig:G1S} for high and low metallicity solar models including macroscopic transport of chemicals and reproducing the lithium depletion observed in the Sun. These aspects must before be taken into account by determining the actual density and pressure values in the solar convective zone from helioseismic data. Moreover, by using various models with transport prescriptions, we also use different helium mass fractions in the convective envelope in the inversion, providing therefore a robust analysis regarding the third term in Eq. \ref{eq:Gamma1Eos}.

\section{Inversions of first adiabatic exponent}\label{Sec:Gamma1}

As shown in Fig \ref{Fig:G1S}, the higher layers of the convective envelope, around $0.95\rm{R}_{\odot}$, are strongly impacted by the EOS and helium ionization. Disentangling the effects of the metal mass fraction in the solar envelope will be extremely difficult. Therefore, we chose to cut the domain in two zones, above $0.91\rm{R}_{\odot}$, the $\Gamma_{1}$ profile will be used to constrain the helium mass fraction, $Y$, whereas the lower part of the domain, below $0.91\rm{R}_{\odot}$ should not bear any strong signal of helium ionization and should be more efficient in isolating the effects of the metallicity. We also see that the slope of the $\Gamma_{1}$ profiles between models computed with FreeEOS and those computed with SAHA-S are at odds around $0.87R_{\odot}$. This plays an important role for the analysis of the solar data as the final accepted Z value might be affected by the choice of the EOS. In what follows we will determine a robust approach to determine Z while minimizing such effects.  

By looking at $\Gamma_{1}$ inversions in the lower parts of the convective zone, we are trying to determine very small corrections, of the order of a few $10^{-4}$. They will be influenced by multiple effects: trade-off parameters of the inversions, surface effects, cross-term contributions. Some can be tested, but will depend on the dataset used in the inversion procedure. Therefore, calibrating the overall procedure with artificial data is as important as carrying out the actual inversion using the observations. Moreover, a measure of the quality of the inversion can be made by verifying that the overall reconstruction of the seismic solar profile is consistent for various reference models. Should large model-dependencies remain, the results would not necessarily be robust and the trade-off parameters could be suboptimal.  

\subsection{The inversion strategy for metallicity determination}\label{Sec:Inversions}

The determination of the metallicity is carried out using the dependencies of $\Gamma_{1}$ on the chemical composition of the envelope. The first adiabatic exponent is a function of four thermodynamic coordinates, namely
\begin{align}
\Gamma_{1}=\Gamma_{1}(\rho,P,Y,Z).
\end{align}
Physically, the most interesting dependencies in the context of this study are those linked with the chemical composition. These will be physically resulting from the ionizations of helium, and those of the heavy elements at different temperatures in the solar envelope (see \citet{Baturin2022} for a dicussion and illustrations). As shown in the left panel Fig. \ref{Fig:G1S} and explicitly visible in Eq. \ref{eq:Gamma1Eos}, a first issue with helioseismic determinations of the solar metallicity is the dependency of the method on the EOS. The only solution to the problem is to actually test the method for various equations of state available to solar modellers. In our case, this is done by using the FreeEOS \citep{Irwin} and two versions of SAHA-S EOS \citep{Gryaznov2004,Gryaznov2006,gryaznov2013,Baturin2013} (see also the SAHA-S
web-site\footnote{\href{http://crydee.sai.msu.ru/SAHA-S_EOS}{http://crydee.sai.msu.ru/SAHA-S\underline{\hphantom{a}}EOS}}).  in the modelling. The configuration used for FreeEOS is the recommended form reproducing the behaviour of the OPAL EOS \citep{Rogers2002} and adopted in the latest generations of Standard Solar Models \citep{Vinyoles2017,Magg2022}. 

\begin{figure*}
	\centering
		\includegraphics[trim= 0 0 0 0, clip, width=11cm]{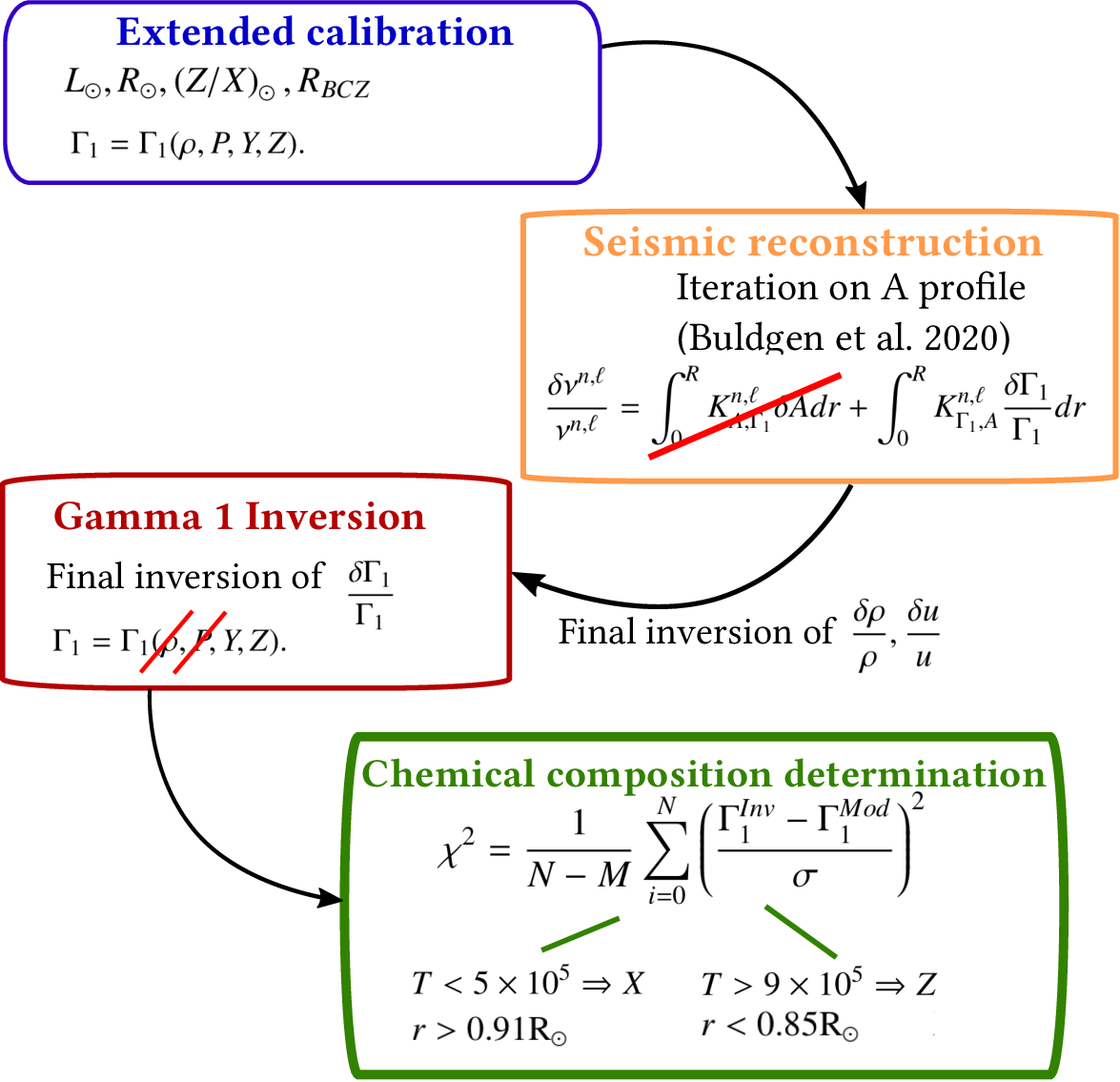}
	\caption{Details of the fitting approach used to determine the chemical composition of the solar envelope. The four-steps process is tailored to eliminate cross-term contributions at each step and maximize the accuracy while the last step computed a reduced $\chi^{2}$ value for each subdomain (see text).}
		\label{Fig:InvMethod}
\end{figure*}

The other uncertainties regarding thermodynamic properties of the convective envelope are taken into account by using inversions for the density, the entropy proxy and squared isothermal sound speed, u=P/$\rho$. These serve to determine the solar conditions as accurately as possible and act as sanity checks for the accuracy and precision of the overall method. When working on actual solar data, to ensure maximal accuracy, we actually use the seismic models as the reference models for the final entropy, isothermal sound speed, and density inversions. The whole procedure for the metallicity inversion is summarized in Fig. \ref{Fig:InvMethod}. The first step is the extended solar calibration (blue box) that provides initial conditions for the second step (orange box): the seismic reconstruction procedure of \citet{Buldgen2020}, that damps the cross-term from the Ledoux discriminant in the variational equation and lead to an accurate depiction of solar thermodynamical conditions. The third step consists in inversions of density (using the ($\rho$, $\Gamma_{1}$ kernels)), squared isothermal sound speed $(u=P/\rho)$ (using the ($u$, $\Gamma_{1}$ kernels) and $\Gamma_{1}$ (using the (A, $\Gamma_{1}$ kernels) to determine the final thermodynamical coordinates (red box) which are then provided to the EOS routine.

This leads to the fourth step of the inference procedure (green box), during which a scan in X and Z is carried out to determine the optimal composition of the solar envelope, using the seismic thermodynamical coordinates determined at the previous step. In this study, we considered X values within $\left[0.70,\; 0.764 \right]$  with a step of $0.001$ and Z values within $\left[ 0.01,\;0.0176 \right]$, with a step of $0.0004$. Therefore a grid of 64 by 20 values is considered for X and Z when reconstructing the $\Gamma_{1}$ profile. The normalized $\chi^{2}$ is computed using the $\Gamma_{1}$ values from the last inversion, with N the number of poins, M is the number of free parameters, here 2 (X and Z), and $\sigma$ is the 1 $\sigma$ uncertainty on the $\Gamma_{1}$ from the inversion.
\\
The domain of the solar convective envelope is separated in two zones. The first one, above $T\approx 9\times 10^{5} K$, is dedicated to the determination of the metallicity. The second one, below $T\approx 5\times 10^{5} K$ is dedicated to the determination of the hydrogen abundance. Due to the much higher abundance of helium in the solar envelope, the effects of the ionization of helium and hydrogen will largely dominate the properties of $\Gamma_{1}$ at lower temperature, but not in the lower part of the CZ \citep[see][for an illustration and a discussion]{Baturin2022}.

\section{Convective envelope composition}\label{Sec:XYZ}

\subsection{Artificial data}\label{Sec:HH}

A first test to determine the robustness of the method is to carry out a full analysis on artificial data. This is done by considering one of the seismic models as a reference model in the procedure while using an evolutionary model as the target. We consider three cases. 

\begin{itemize}
\item \textbf{HH1:} M1 plays the role of the target and A2 is the reference, thus both effects of abundances and EOS are considered.
\item \textbf{HH2:} A1 is the target and A3 is the reference. In this case no metallicity correction should be found and the corrections are purely EOS effects. 
\item \textbf{HH3:} M2 is the target and A2 is the reference, only chemical composition effects are considered as the EOS is the same for both models. 
\end{itemize}

The dataset considered for the tests is exactly the same as the actual solar one, with the same uncertainties on individual modes. Therefore the propagation of uncertainties, calibration of the trade-off parameters and effects of trade-off, surface corrections,... should be as similar as possible with respect the procedure used for actual solar data. We voluntarily chose high-Z models as targets to simulate the effect of a high-Z Sun, following the work of \citet{Magg2022}, to see whether this could be recovered from the data.  

\begin{figure*}
	\centering
		\includegraphics[trim= 0 0 0 0, clip, width=17cm]{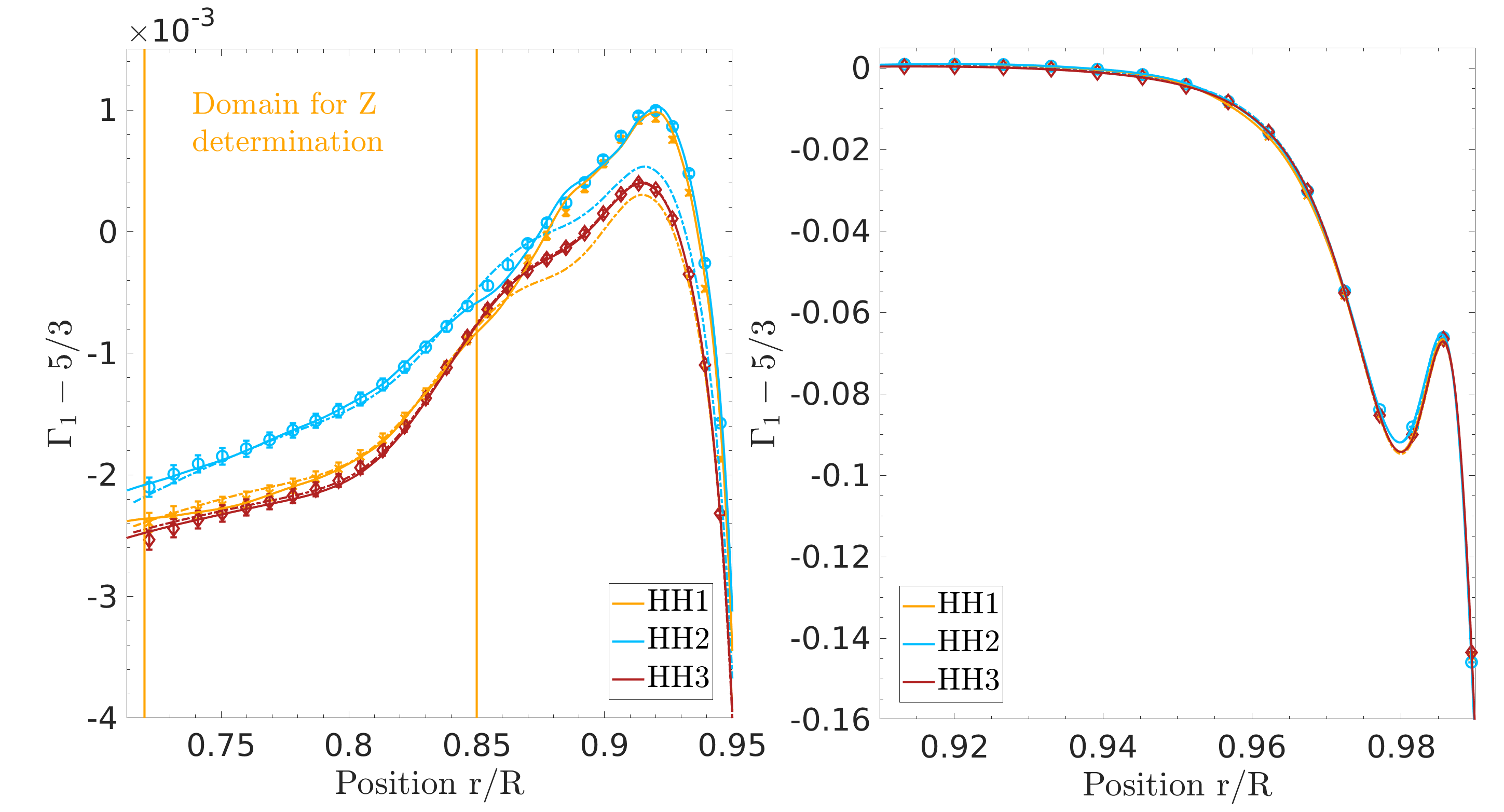}
	\caption{Inversions of the $\Gamma_{1}$ profile as a function of normalized radius for the artificial data (HH cases). Each panel illustrates a subdomain of the method: left panel for the Z determination, indicated by the orange vertica lines, right panel for the X determination.}
		\label{Fig:InvHHG1}
\end{figure*}

Before entering in the details of the analysis of Fig. \ref{Fig:InvHHG1}, we provide some additional details on the technicalities of the inversion. The full domain of the inversion spans from $\approx 0.72R_{\odot}$ up to $\approx 0.985R_{\odot}$. Below this lowest value, the inversion might be significantly affected by the effects of the boundary of the convective zone, $0.72$ can actually already be considered quite low and explains why a high trade-off parameter for the cross-term integral has been considered. Above $0.985$, the inversion starts to be affected by the outer boundary conditions, the averaging kernels, despite showing good localization, show sometimes sharp deviations. Moveover, the inversion is limited by the availability of other coordinates (namely $u=\frac{P}{\rho}$ and $\rho$) which are more difficult to localize at such high radii with the considered dataset. Robustness with respect to surface effects, mass conservation constraint in the model might also play a role and we chose to be conservative. 

The overall domain is then subdivided in two subdomains. The first one, at lower radii and higher temperatures, will span from $\approx 0.72R_{\odot}$ up to $\approx 0.85R_{\odot}$. It counts about 17 points with $\Gamma_{1}$ inversion values. Due to the higher temperatures, all of H and He are ionized and while both elements will contribute to a large fraction of the electrons in these regions, the dips in $\Gamma_{1}$ will be mostly affected by the partial ionization of metals. The effect is illustrated in \citet{Baturin2022} Fig. 5, upper panel, where one can see that the region of interest will be mostly influenced by oxygen and slightly influenced by carbon. Therefore, the Z inversion performed here will mostly validate the abundance of oxygen, although the details of this should be confirmed from an analysis using the approach of \citet{Baturin2022}. Due to the differences in $\Gamma_{1}$ between FreeEOS and SAHA-S between $\approx 0.85R_{\odot}$ and $\approx 0.91R_{\odot}$, we chose first to neglect this region of the inversion (except in Sec. \ref{Sec:EOS}). While this might appear as a strong hypothesis, it is merely a choice of using 17 constraints in a region where all equations of state agree to determine one free parameter of the thermodynamical properties of the envelope, namely Z. The underlying hypothesis of the method being that if FreeEOS and SAHA-S provide the same $\Gamma_{1}$ values for the same coordinates, then the physics of the EOS must be robust\footnote{We mention that this actually does not influence the conclusions of the study on solar data, rather it makes the procedure more difficult for FreeEOS of the actual solar data analysis, but does not affect the conclusion that if the Sun was high-Z, we should pick up the signal in the $\Gamma_{1}$ profile.}.  

In this work, we will focus on a global determination of Z. If the hypothesis that Z dominates largely in the properties of $\Gamma_{1}$ in the first subdomain is valid, a scan in X and Z as discussed above should provide an almost horizontal valley of optimal values of Z for various values of X in an X-Z $\chi^{2}$ map. This will be verified below.

The second subdomain is considered above $\approx 0.91R_{\odot}$ up to  $\approx 0.985R_{\odot}$. It is used to constrain the X value in the convective envelope. This is similar to what has been done in the past litterature \citep[e.g.][]{Vorontsov91,BasuYSun,RichardY}. The choice of going for X instead of Y does not affect the conclusions, as the couple X and Z will determine a Y value and in these low temperature regions, the effect of the metals is almost insignificant with respect to the uncertainties due to the EOS, surface effects and inaccuracies in the thermodynamical coordinates. Again this hypothesis can be checked when drawing the $\chi^{2}$ map as the optimal solution should appear as a vertical valley in a X-Z plane. This will be again verified below. 

The results are illustrated in Fig. \ref{Fig:InvHHG1} for the $\Gamma_{1}$ inversion. Comparing the full lines and the inversion results, we can see that the inversion reproduces quite nicely the actual behaviour of the profile. This means that the trade-off parameters have been well adjusted. Now if we look at all cases, we see that in the left panel, the reconstructions have been able to grasp most of the features in the fitted areas. In each case where a high-Z model was a target  (green and red symbols), the inversion managed to pick it up quite nicely, while the discrepancies between FreeEOS and SAHA-S between $\approx 0.85R_{\odot}$ and $\approx 0.91R_{\odot}$ are clearly seen for HH1. Similarly when both models were of low-Z values but still exhibited significant differences due to their differing equations of state, the inversion managed to recover the proper range of metallicity. The same can be seen in the right panel for the helium ionization zones.

\begin{figure*}
	\centering
		\includegraphics[trim= 0 0 0 0, clip, width=16cm]{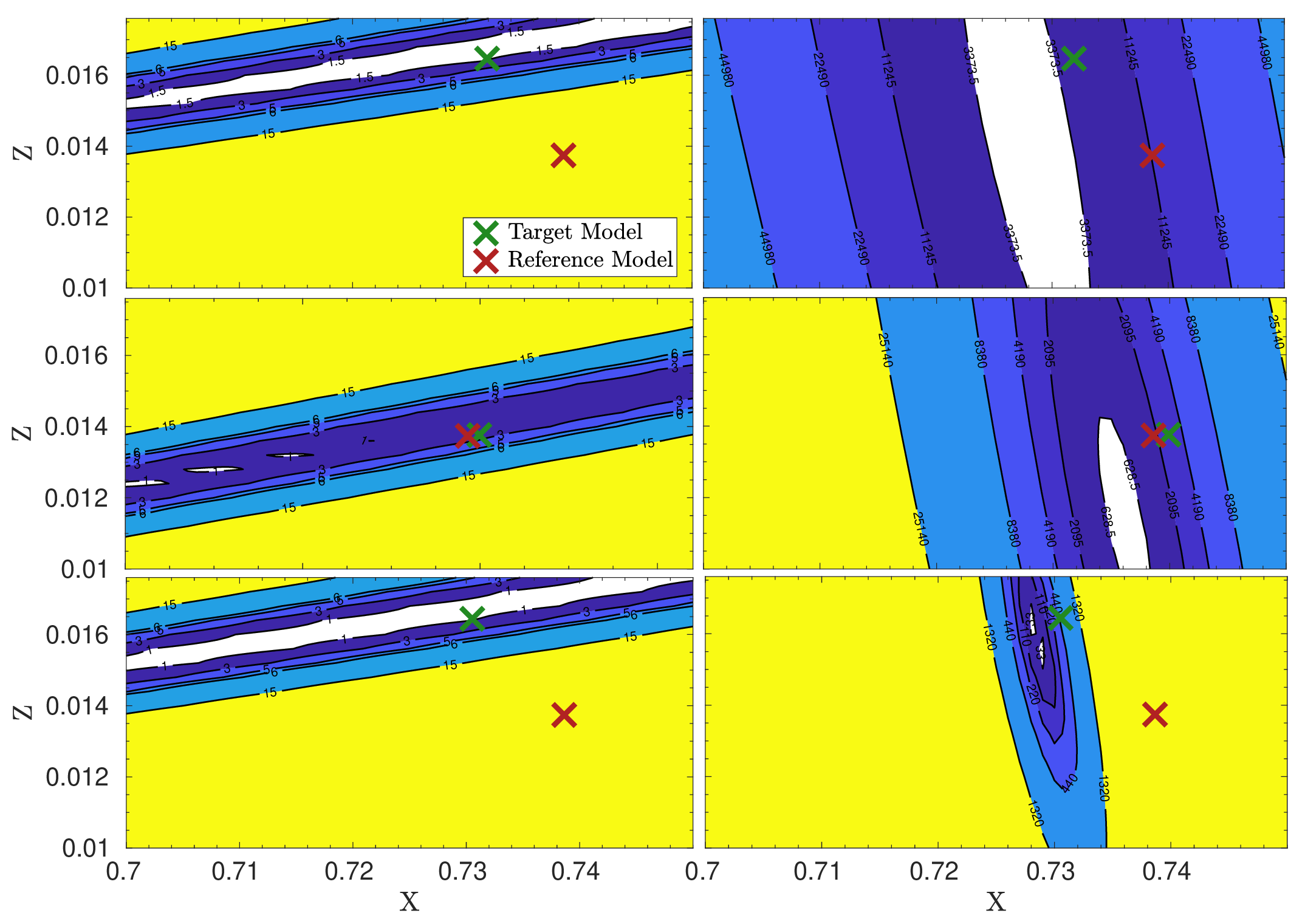}
	\caption{$\chi^{2}$ map of the X and Z scan for the artificial data (HH exercises). The green and red crosses indicate the target and reference model, respectively. The upper panels are for HH1, middle panels for HH2 and lower panels for HH3. The left panels are associated with the high-T subdomain for the Z determination, while the right panels are associated with the low-T subdomain used to determine X.}
		\label{Fig:MapHHNew}
\end{figure*}

The $\chi^{2}$ maps are illustrated in Fig \ref{Fig:MapHHNew}, with the left panel of each figure showing the results for the high-T subdomain used to determine Z and the right panel showing the results for the low-T subdomain used to determine X. We can see that the inversion actually provides a quite accurate estimate of both X and Z in all cases. While the Z inversion is not perfectly horizontal, a clear favoured region can be outlined and using the information on X from the right panel to a favoured Z in the left panel helps further refining the information on Z. Here, we chose to limit the valley in X to the regions where the $\chi^{2}$ was below $1.5\times \chi^{2}_{\rm{Min}}$, with $\chi^{2}_{\rm{Min}}$ being the lowest $\chi^{2}$ found for all values of X and Z used in the scan at low temperatures. This interval is then used to constrain the Z range in the valley, where the criterion is either to have $\chi^{2}<1$ whenever it is reached in an extended range, or $\chi^{2}<1.5\times \chi^{2}_{\rm{Min}}$ when the former criterion is not satisfied. In all cases this leads to a determined metallicity in agreement with the target value. We can see that this approach makes the technique less dependent on the details of the equation of state at high temperatures. In practice the change in reduced $\chi^{2}$ value induced by using the information of X (using the value of X with lowest $\chi^{2}$ to constrain Z) from the lower temperature is minimal and below $1$, meaning that the fit remains consistent. In all cases this approach provides X and Z values in good agreement with the actual values of the models. For HH1, the X interval found is $\left[ 0.724,\;0.733 \right]$ and the Z interval is $\left[0.0160,\; 0.0170 \right]$ with the solution being $\rm{X}=0.732$, $\rm{Z}=0.01647$. For HH2, the X interval found is $\left[ 0.734,\;0.738 \right]$ and the Z interval is $\left[0.0136,\; 0.0142 \right]$ with the solution being $\rm{X}=0.7386$, $\rm{Z}=0.01374$. Finally, for HH3, the X interval found is $\left[ 0.727,\;0.729 \right]$ and the Z interval is $\left[0.0163,\; 0.0169 \right]$ with the solution being $\rm{X}=0.73056$, $\rm{Z}=0.01644$.  While slight biases can be observed for X, a clear result is that a low-Z Sun cannot be mistaken with a high Z Sun using current modern equations of state and that the provided interval predicts the correct value. Actual reduced $\chi^{2}$ differences between high-Z and low-Z models range between a factor 5 and 8 for the scan in metallicity. The high $\chi^{2}$ values for the X inversion are found in the cases for which the equation of state is different between the target and the reference models. In these cases, the $\chi^{2}$ values always keep values of about $1000$ or few hundreds at best. This is due to large discrepancies between the equations of state, amplified by the very small uncertainties of the SOLA inversion in regions where the method might actually be less robust (for reasons mentioned above). The color scale in Fig \ref{Fig:MapHHNew} is as follows: for the high-T subdomain used to determine Z: white corresponds to $\chi^{2}< 1$, successive shades of blue to $\chi^{2}< 3$, $\chi^{2}< 5$, $\chi^{2}< 6$ and yellow corresponds to $\chi^{2}> 15$. For the low-T subdomain used to determine X: white corresponds to $\chi^{2}<1.5\times \chi^{2}_{\rm{Min}}$, successive shades of blue to $\chi^{2}<5\times \chi^{2}_{\rm{Min}}$, $\chi^{2}<6\times \chi^{2}_{\rm{Min}}$, $\chi^{2}<15\times \chi^{2}_{\rm{Min}}$ and yellow corresponds to $\chi^{2}>25\times \chi^{2}_{\rm{Min}}$, with $\chi^{2}_{\rm{Min}}=2249$, for HH1, $419$ for HH2, $22$ for HH3.

\subsection{Solar data}\label{Sec:RealData}

\begin{figure*}
	\centering
		\includegraphics[trim= 0 0 0 0, clip, width=17cm]{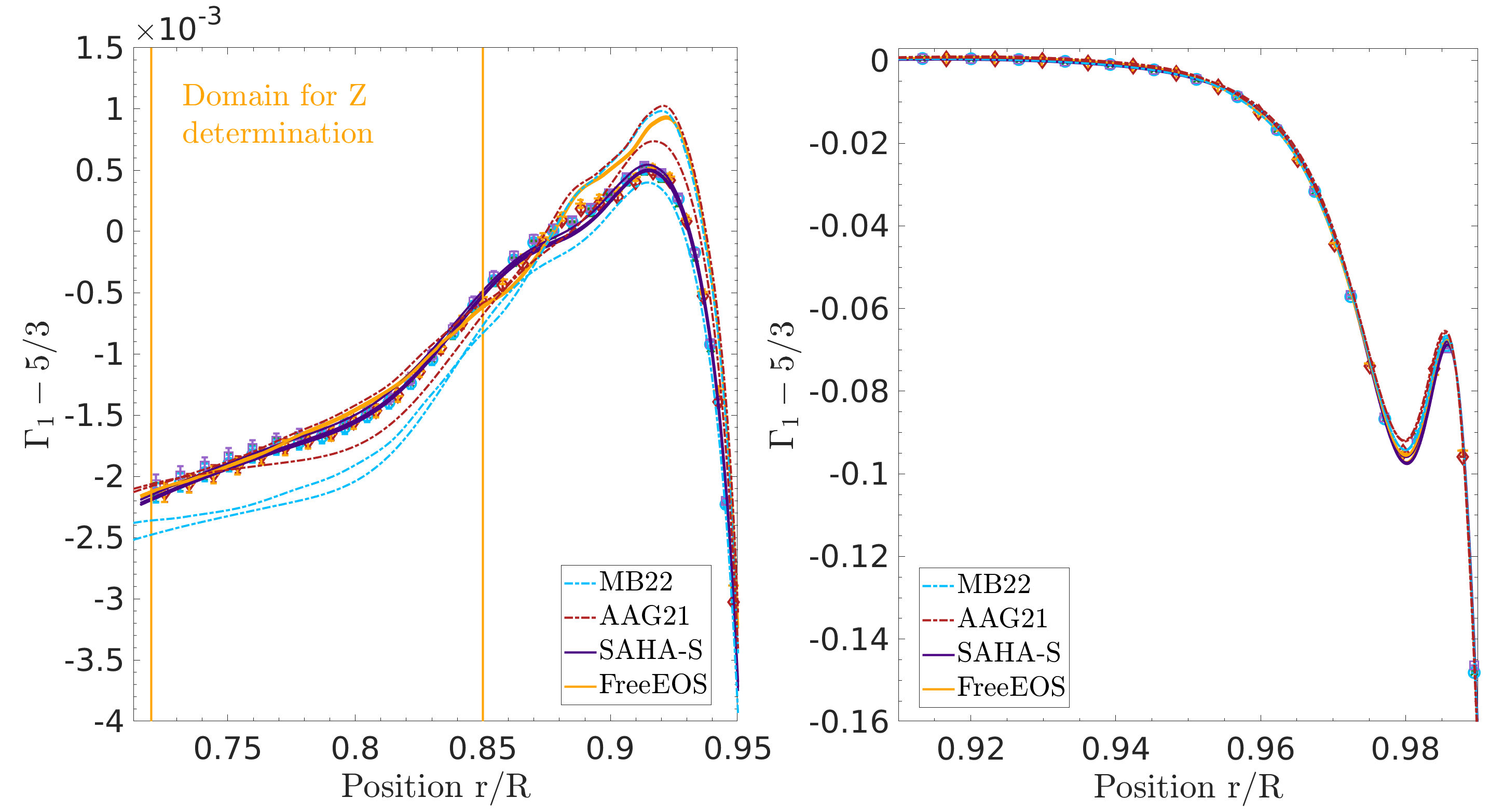}
	\caption{Inversions of the $\Gamma_{1}$ profile as a function of normalized radius determined from actual solar data. Each panel illustrates a subdomain of the method: left panel for the Z determination, indicated by the vertical orange lines, right panel for the X determination. The orange curves are associated with reconstructions using FreeEOS, the purple curves are associated with reconstructions using SAHA-S and the blue and red curves are associated with models M1 and M2 and A1 and A2, respectively.}
		\label{Fig:InvG1SunAll}
\end{figure*}

We start by presenting in Fig. \ref{Fig:InvG1SunAll} the $\Gamma_{1}$ inversion results for each of the 5 seismic models presented in Sect. \ref{Sec:Models}. It appears that the inversions, ploted using the symbols of various colors, are consistent with each other for all models. This gives confidence that the $\Gamma_{1}$ profile has been accurately determined in a model-independent way. 

A quick look at the left panel Fig. \ref{Fig:InvG1SunAll} already shows that the \citet{Magg2022} models, denoted by the blue dashed lines, are at odds with the data in the lower part of the convective zone, while this is not the case or at least less the case of the AAG21 models shown in red. A second observation on the slope of $\Gamma_{1}$ between $0.85R_{\odot}$ and $0.91R_{\odot}$ indicates that solar data seems to strongly favour the SAHA-S equation of state. However a physical explanation for these differences remains to be provided as it could have multiple origins (coulomb correction, abundance of specific individual elements \citep{Trampedach2006}). This observation confirms that the metallicity can be inferred most robustly between $0.72$ and $0.85R_{\odot}$. As mentioned in the previous section, we thus have 17 $\Gamma_{1}$ inversion values for the first subdomain. Tests with the full profile between $0.72$ and $0.91R_{\odot}$ have also been performed in Sect. \ref{Sec:EOS}. 

In the left panel of Fig. \ref{Fig:InvG1SunAll}, the best fit profiles using FreeEOS between $0.72$ and $0.85R_{\odot}$ are provided by the orange curves while those using SAHA-S EOS are provided by the dark blue curves. These models also reproduce very well the solar data up to $0.91R_{\odot}$, while these additional points are not included in the fit yet. For each EOS, the curves obtained through the reconstruction are essentially the same, meaning that the procedure is independent of the reference model and that only the assumed EOS might affect the final result. This effect is however mitigated by limiting the first subdomain between $0.72$ and $0.85R_{\odot}$.

In the right panel of Fig. \ref{Fig:InvG1SunAll}, we take a look at the helium second ionization zone. In this case the situation is reversed, the \citet{Magg2022} models, with their higher helium abundance in the solar convective zone, are in much better agreement then the AAG21 models. This is confirmed using the $\Gamma_{1}$ reconstruction that favour very high helium abundances in the solar convective envelope. Again all inversion points tend to agree with each other, providing confidence in the robustness of the approach. However, the points above $0.98R_{\odot}$ might be influenced by boundary effects, surface effects or slight inaccuracies in the determination of the $\rho$ and $u$ coordinates, as OLA inversions tend to be less accurate at the borders of the domain \citep{Backus1968,Pijpers}. 

Nevertheless, the results can be used to compute a $\chi^{2}$ map for the two subdomains: between $0.72$ and $0.85R_{\odot}$ to constrain Z, and between $0.91$ and $0.985R_{\odot}$ to constrain X. Combining the optimal X and Z found in both subdomains, we get an accurate estimate of both parameters, as demonstrated on artificial data in the previous section. 

\begin{figure*}
	\centering
		\includegraphics[trim= 0 0 0 0, clip, width=16cm]{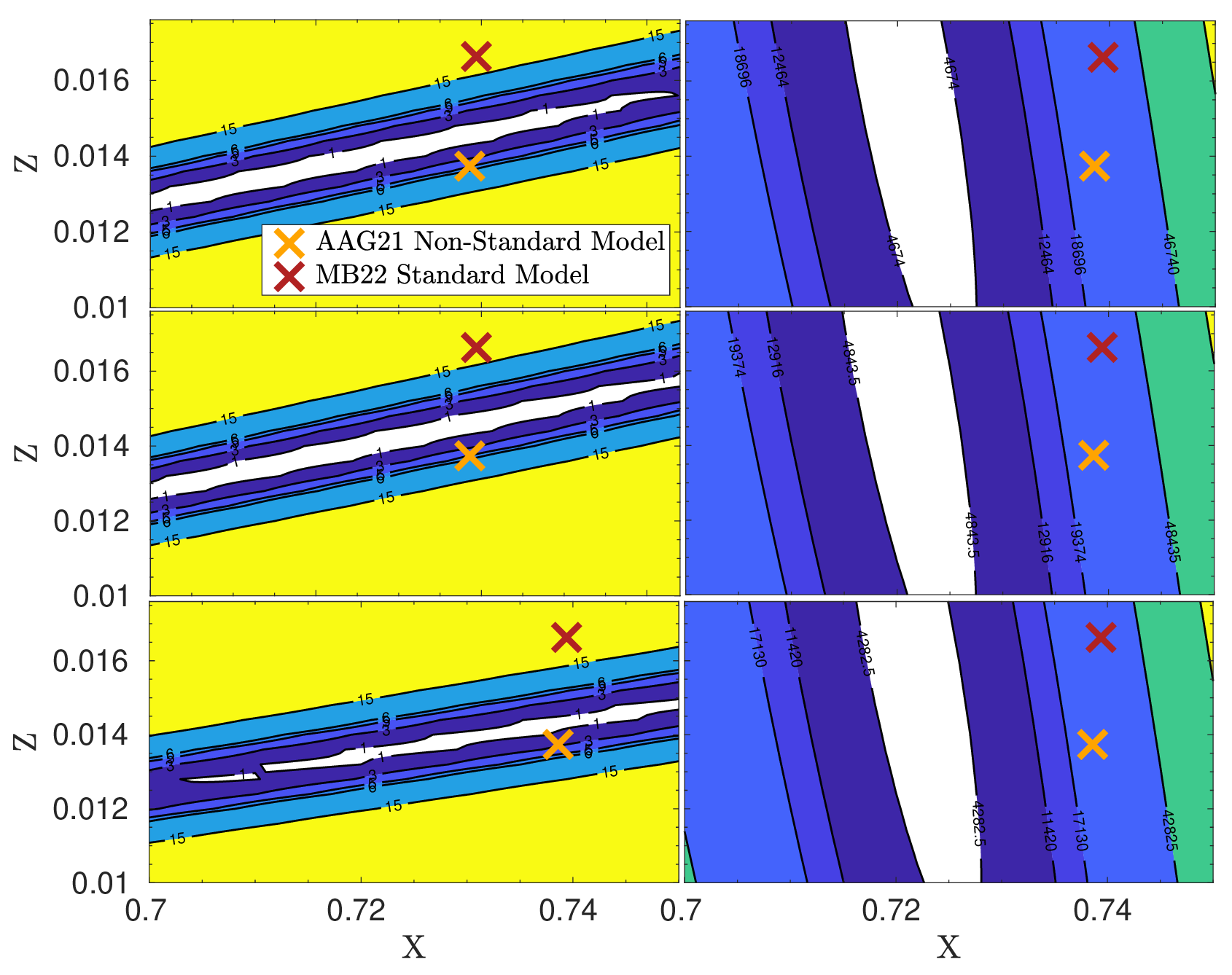}
	\caption{$\chi^{2}$ map of the X and Z scan for solar data, using Model A2 (upper panels), A3 (middle panel) and M2 (lower panels) in the procedure, therefore either SAHA-S v7 or SAHA-S v3. the orange and red crosses indicate the positions of AAG21 model including rotation and magnetic fields and the MB22 standard solar model respectively (values from their paper). The left panel is associated with the high-T subdomain for the Z determination, while the right panel is associated with the low-T subdomain used to determine X.}
		\label{Fig:MapA2A1A3}
\end{figure*}

\begin{figure*}
	\centering
		\includegraphics[trim= 0 0 0 0, clip, width=15cm]{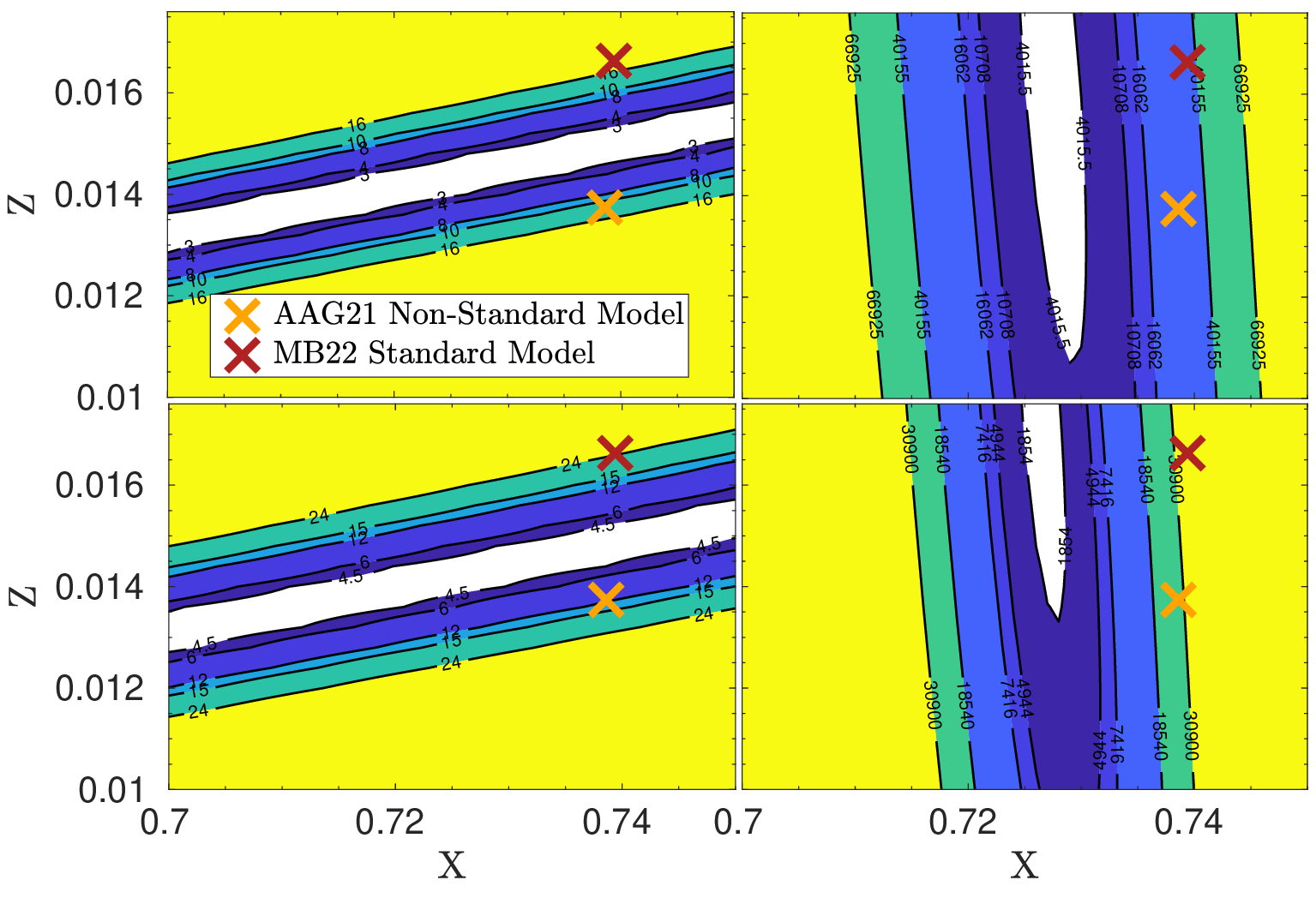}
	\caption{Same as Fig. \ref{Fig:MapA2A1A3} for models A1 (upper panels) and M1 (lower panels), therefore using FreeEOS in the reconstruction. The left panel is associated with the high-T subdomain for the Z determination, the right panel is associated with the low-T subdomain used to determine X.}
		\label{Fig:MapM2M1}
\end{figure*}

The results are illustrated in Figs. \ref{Fig:MapA2A1A3} and \ref{Fig:MapM2M1} where the $\chi^{2}$ maps show that the assumption of separating the domain works quite well. Indeed, the optimal solution in the right panels are almost vertical, indicating that there is little dependence on the metallicity and that, as expected, X dominates the solution (due to its direct impact on Y and thus on the reproduction of the properties of the helium ionization regions). Similarly, at high temperatures, a clear region in Z is outlined as optimal solution. It is also clear that the higher temperature layers bear little to no information on X, as the material is clearly fully ionized. Thus there is no clear trace left in the $\Gamma_{1}$ profile to directly infer X. Therefore, as already seen in the HH exercise, one can use the information on X from lower temperatures to constrain the optimal interval for Z in the observed $\chi^{2}$ valley in the left panels of Figs. \ref{Fig:MapA2A1A3} and \ref{Fig:MapM2M1}. This provides, independently of the equation of state used (but assuming that either FreeEOS or SAHA-S is equally good in representing the plasma in the solar envelope) a final X interval between $\left[0.715,\;0.730\right]$ and a final Z interval between $\left[0.0132,\;0.0148\right]$ for the first dataset studied here. 

The color scale in Fig \ref{Fig:MapA2A1A3}, using SAHA-S (v3 or v7) in the analysis, is the following: for the high-T subdomain: white corresponds to $\chi^{2}<1$, successive shades of blue to $\chi^{2}<3$, $\chi^{2}<5$, $\chi^{2}<6$ and yellow corresponds to $\chi^{2}>15$; for the low-T subdomain used to determine X: white corresponds to $\chi^{2}<1.5\times \chi^{2}_{\rm{Min}}$, successive shades of blue to $\chi^{2}<4\times \chi^{2}_{\rm{Min}}$, $\chi^{2}<6\times \chi^{2}_{\rm{Min}}$, $\chi^{2}<15\times \chi^{2}_{\rm{Min}}$ and yellow corresponds to $\chi^{2}>25\times \chi^{2}_{\rm{Min}}$, $\chi^{2}_{\rm{Min}}=3115$ for model A2, $3229$ for model A3 and $2855$ for model M2.

The color scale in Figs \ref{Fig:MapM2M1} where FreeEOS is used in the analysis, is the following: for the high-T subdomain: white corresponds to $\chi^{2}<1.5\times \chi^{2}_{\rm{Min}}$, successive shades of blue to $\chi^{2}<2\times \chi^{2}_{\rm{Min}}$, $\chi^{2}<4\times \chi^{2}_{\rm{Min}}$, $\chi^{2}<6\times \chi^{2}_{\rm{Min}}$ and yellow corresponds to $\chi^{2}>8\times \chi^{2}_{\rm{Min}}$, with $\chi^{2}_{\rm{Min}}=1.8$ starting from model A1 and $3.14$ starting from model M1; for the low-T subdomain used to determine X: white corresponds to $\chi^{2}<1.5\times \chi^{2}_{\rm{Min}}$, successive shades of blue to $\chi^{2}<4\times \chi^{2}_{\rm{Min}}$, $\chi^{2}<6\times \chi^{2}_{\rm{Min}}$, $\chi^{2}<15\times \chi^{2}_{\rm{Min}}$ and yellow corresponds to $\chi^{2}>25\times \chi^{2}_{\rm{Min}}$, with $\chi^{2}_{\rm{Min}}=2676$ starting from model A1 and $1236$ starting from model M1. The explanation for the $\chi^{2}$ values remaining always above $1$ in the high-T domain is due to the use of the FreeEOS equation of state. This means that SAHA-S tends to provide overall better fits to solar data than FreeEOS at higher temperatures. To the contrary, FreeEOS tends to provide a better fit for the low T domain overall, although none of the $\chi^{2}$ values match as well as in HH3, suggesting inaccuracies in the EOS in these low T regimes and perhaps systematics in the inversion.

\subsection{Assuming the equation of state known}\label{Sec:EOS}

As mentioned above, SAHA-S EOS provides a much better agreement with solar data than FreeEOS. Therefore, it is worth investigating what conclusions we can draw using the full set of $\Gamma_{1}$ points and assuming SAHA-S as the equation of state describing the solar material. We check whether in these conditions one may infer individual elements abundances, as carried out in \citet{Baturin2022}. To do so, we attempt at reconstructing $\Gamma_{1}$ using either the AGSS09 or the MB22 abundances, using X and Z as free parameters, starting from both model M1 and A2. 

\begin{figure*}
	\centering
		\includegraphics[trim= 0 0 0 0, clip, width=14cm]{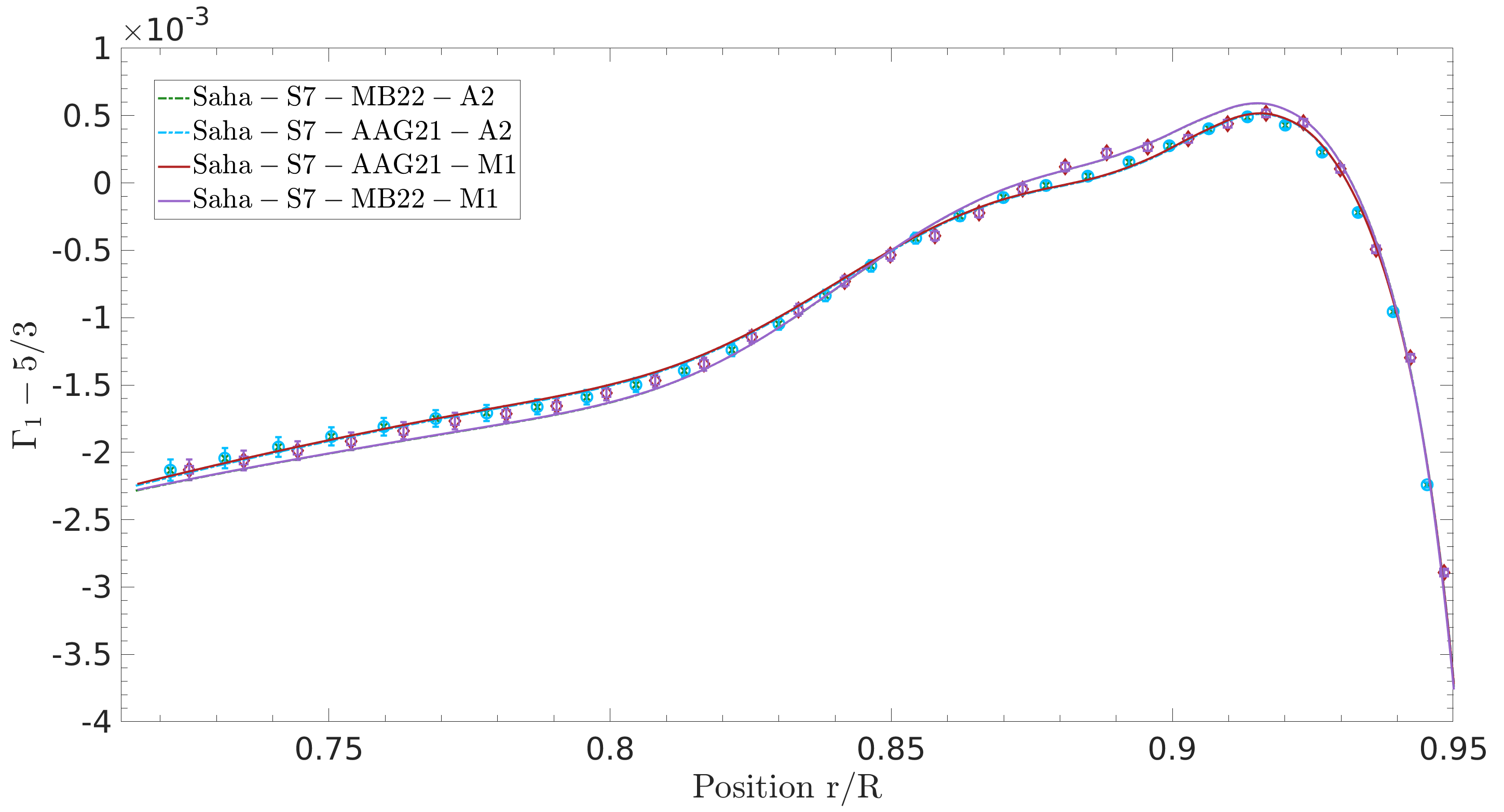}
	\caption{Inversions of the $\Gamma_{1}$ profile as a function of normalized radius determined from solar data, assuming SAHA-S v7 as the equation of state and changing the ratios of individual elements in the table (from MB22 to AAG21). Two reference models are used in the procedure, model M1 and A2, the results of the reconstructions are plotted in various colors. The red curve overplots the light blue one and the purple curve overplot the green one. Each symbol depicts the $\Gamma_{1}$ inversion results for the associated model (M1 and A2).}
		\label{Fig:G1Abund}
\end{figure*}

\begin{figure*}
	\centering
		\includegraphics[trim= 0 0 0 0, clip, width=15cm]{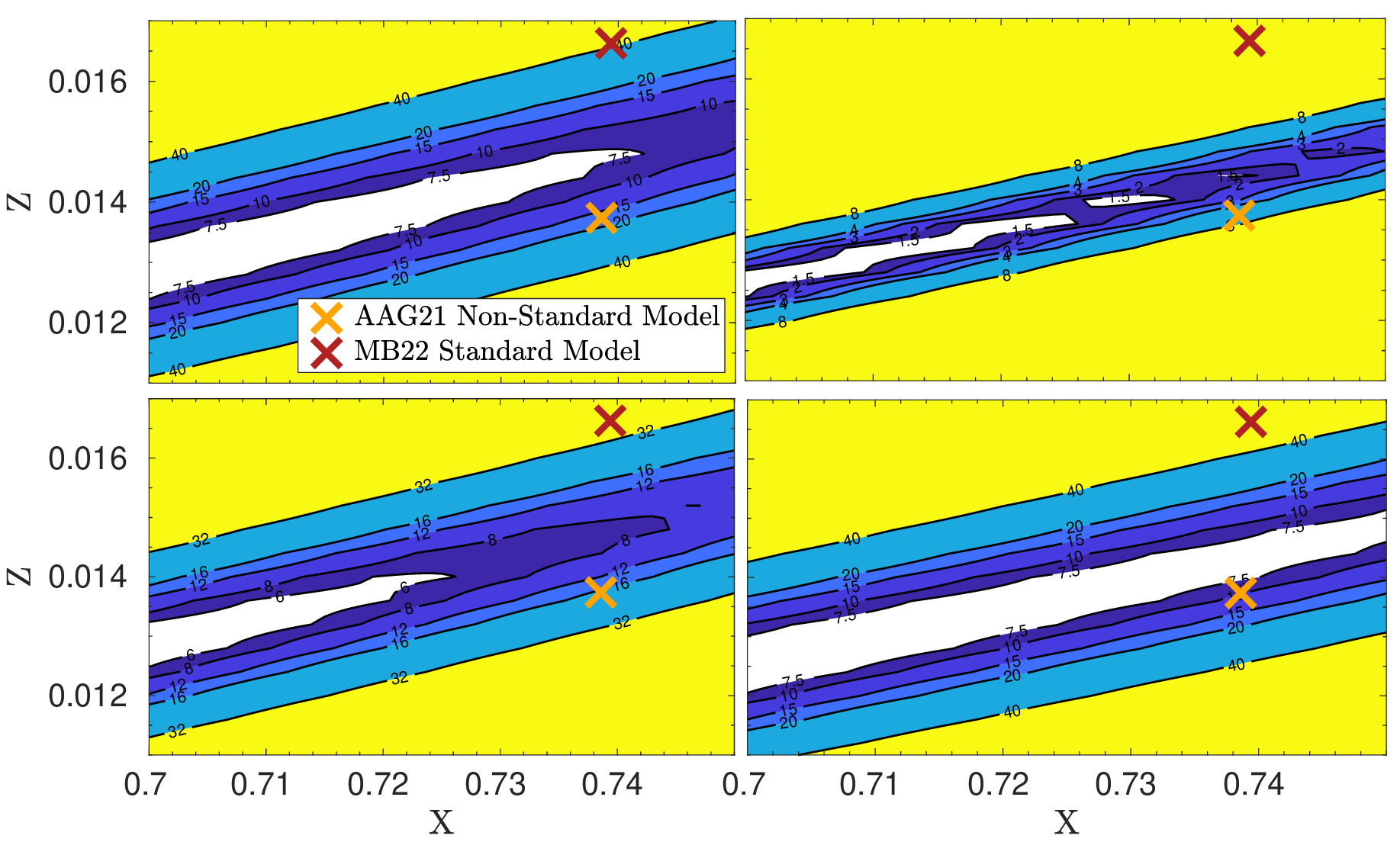}
	\caption{$\chi^{2}$ map of the X and Z scan for solar data fitting $\Gamma_{1}$ between $0.72R_{\odot}$ and $0.91R_{\odot}$,  the orange and red crosses indicate the positions of AAG21 model including rotation and magnetic fields and the MB22 standard solar model (MB22-Phot in table \ref{tabModels}) respectively (values from their paper). Upper panels are using model A2, lower panels using model M1/ Left panels are using MB22 individual elements, right panels using AAG21 individual elements.}
		\label{Fig:MapFullDomain}
\end{figure*}

The results are illustrated in Fig \ref{Fig:G1Abund} for the $\Gamma_{1}$ profiles. We can see that a small difference can be made between AGSS09 and MB22 individual ratios of elements, but not at a high level of significance. However, a low Z value is still strongly favoured in the $\chi^{2}$ map illustrated in Fig \ref{Fig:MapFullDomain}. In both cases, a Z value in line with AAG21 is favoured, while the low Z valley seems to be a bit wider if AGSS09 individual ratios of elements are assumed. This implies that while the $\Gamma_{1}$ profile favours a low Z value, in line with the AAG21 photospheric abundances, it might be difficult to disentangle the contribution of individual elements, although trying to pick some dominant trends regarding Oxygen would still be worth attempting. 

The color scale in Fig \ref{Fig:MapFullDomain}, using SAHA-S in the analysis, is the following: white corresponds to $\chi^{2}<1.5\times \chi^{2}_{\rm{Min}}$, successive shades of blue to $\chi^{2}<2\times \chi^{2}_{\rm{Min}}$, $\chi^{2}<3\times \chi^{2}_{\rm{Min}}$, $\chi^{2}<4\times \chi^{2}_{\rm{Min}}$ and yellow corresponds to $\chi^{2}>8\times \chi^{2}_{\rm{Min}}$, with$\chi^{2}_{\rm{Min}}=0.86$ starting from model A2 and the AAG21 abundances (upper right panel), $\chi^{2}_{\rm{Min}}=4.9$ for model A2 with either MB22 (upper left panel) and $\chi^{2}_{\rm{Min}}=4.6$ and $4.2$ starting from model M1 with the MB22 (lower right panel) and AAG21 abundances (lower left panel), respectively.

\subsection{Impact of the dataset}\label{Sec:Data}

\begin{figure*}
	\centering
		\includegraphics[trim= 0 0 0 0, clip, width=14cm]{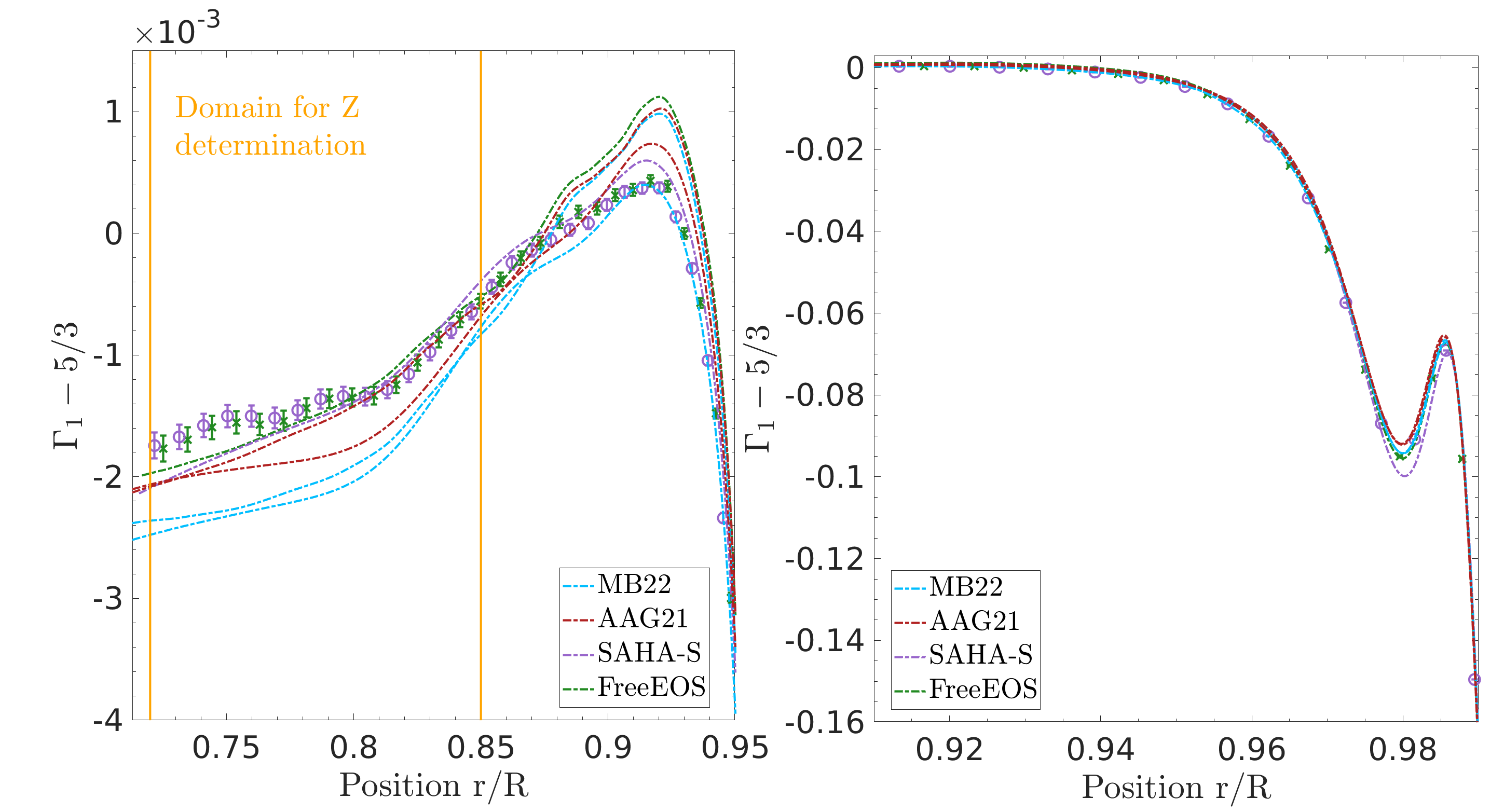}
	\caption{Inversions of the $\Gamma_{1}$ profile as a function of normalized radius in the solar enveloped, determined using the \citet{Larson2015} dataset. Each panel illustrates a subdomain of the method: left panel for the Z determination, indicated by the orange vertical lines, right panel for the X determination. The green curve and symbols are associated with Model M1 (FreeEOS), whereas the purple curve and symbols are associated with model A2 (SAHA-S v7). The red and blue lines illustrate models A1 and A2 and M1 and M2 respectively.}
		\label{Fig:G1Schou}
\end{figure*}

A last test has been performed using more recent MDI data from \citet{Larson2015} to determine whether the trends picked in the previous sections are spurious or not. For this, the whole calibration procedure has been carried out again from the start. The results of the $\Gamma_{1}$ inversions for models M1 and A2 are illustrated in Fig \ref{Fig:G1Schou}. Again, a clear rejection of the high Z solution of \citet{Magg2022} can be observed in the left panel of Fig \ref{Fig:G1Schou}. 

In this case the reconstruction procedure has struggled a bit more to reproduce the $\Gamma_{1}$ values in the lower part of the convective envelope. SAHA-S EOS is still favoured between $0.85R_{\odot}$ and $0.91R_{\odot}$ but even SAHA-S models struggles at reconstructing the profile. Nevertheless, as shown by the blue and red curves, low Z models (in red) are strongly favoured over high Z models (in blue) for the lower part of the CZ (left panel), while a high Y value is still favoured (right panel). The effects are confirmed for both test cases using Model M1 and Model A2.

\begin{figure*}
	\centering
		\includegraphics[trim= 0 0 0 0, clip, width=15cm]{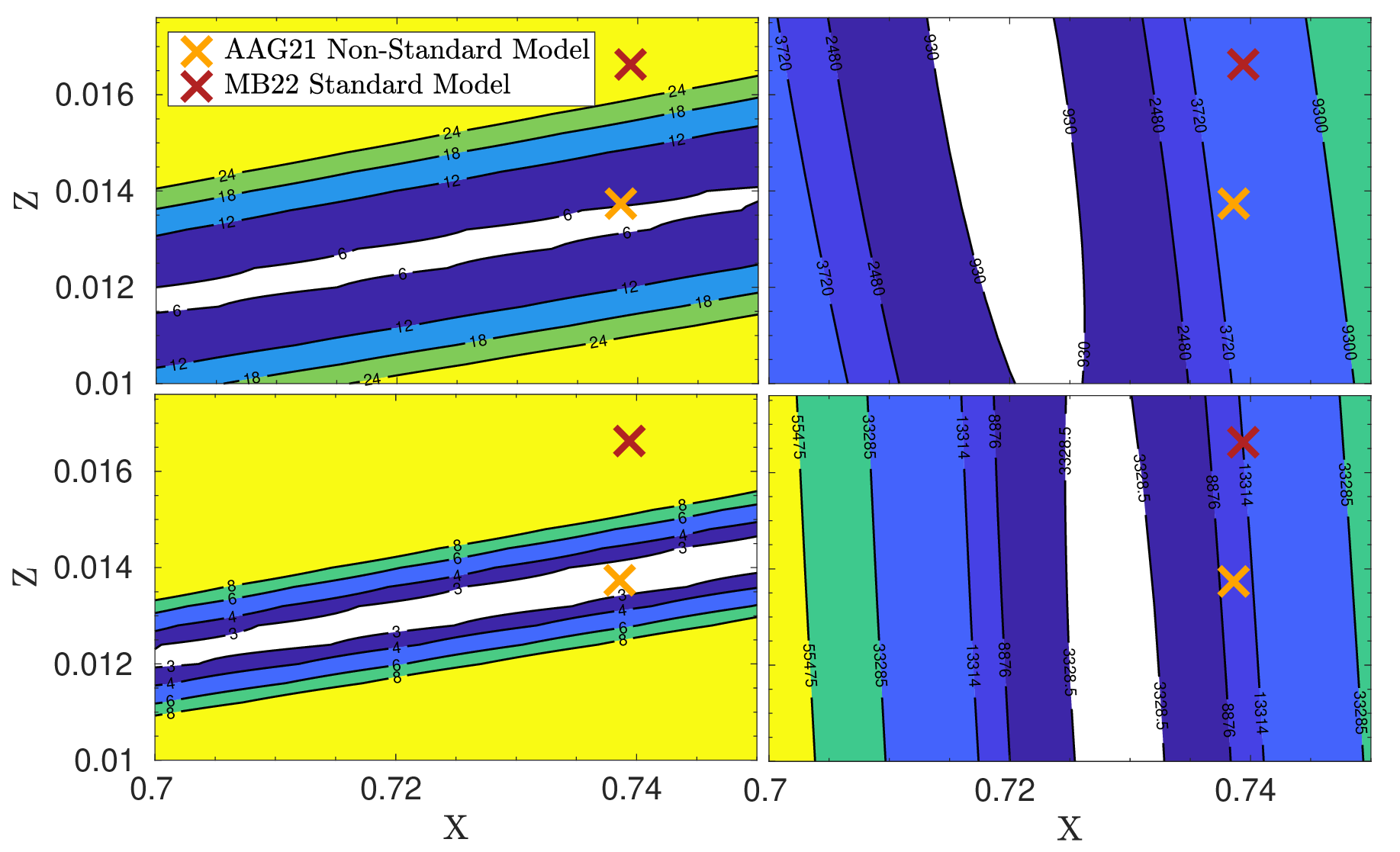}
	\caption{$\chi^{2}$ map of the X and Z scan for solar data \citep{Larson2015}, using Model A2 in the procedure, fitting $\Gamma_{1}$ between $0.72R_{\odot}$ and $0.85R_{\odot}$, the orange and red crosses indicate the positions of AAG21 model including rotation and magnetic fields and the MB22 standard solar model respectively (values from their paper).}
		\label{Fig:MapSchou}
\end{figure*}

The $\chi^{2}$ maps illustrated in Fig. \ref{Fig:MapSchou} further confirm these trends, with low Z values strongly favoured over high Z values. However in this case, a slightly lower value of Z seems to be favoured. This might be similar to the trend seen \citet{Vorontsov13,VorontsovSolarEnv2014}. We can however see that this time, the valley in X is almost perfectly vertical, with a slightly higher Y interval than in previous sections. The overall fit in Z is somewhat more difficult given the strong deviations in the first few points in the deep convective zone (around $0.75R_{\odot}$). In this case again, a factor between 6 and 10 is found between the optimal solution at low Z and the MB22 Standard Solar Model (SSM). While the AAG21 non-standard model performs better, it still unable to reach the low X regime favoured by the inversion. 

The color scale in Fig \ref{Fig:MapSchou}, using SAHA-S in the analysis, is the following: for the high-T subdomain: white corresponds to $\chi^{2}<1.5\times \chi^{2}_{\rm{Min}}$, successive shades of blue to $\chi^{2}<2\times \chi^{2}_{\rm{Min}}$, $\chi^{2}<3\times \chi^{2}_{\rm{Min}}$ and yellow corresponds to $\chi^{2}<4\times \chi^{2}_{\rm{Min}}$, with $\chi^{2}_{\rm{Min}}=5.6$ starting from model A2 using SAHA-S v7 and $\chi^{2}_{\rm{Min}}=2.2$ starting from model M1 and using FreeEOS; for the low-T subdomain used to determine X: white corresponds to $\chi^{2}<1.5\times \chi^{2}_{\rm{Min}}$, successive shades of blue to $\chi^{2}<4\times \chi^{2}_{\rm{Min}}$, $\chi^{2}<6\times \chi^{2}_{\rm{Min}}$, $\chi^{2}<15\times \chi^{2}_{\rm{Min}}$ and yellow corresponds to $\chi^{2}>25\times \chi^{2}_{\rm{Min}}$, with $\chi^{2}_{\rm{Min}}=620$, starting from model A2 and using SAHA-S v7 and $\chi^{2}_{\rm{Min}}=2219$ starting from model M1 and using FreeEOS. The trend observed here is the opposite of what was seen before, with SAHA-S being favoured at low T while FreeEOS is favoured between 0.72 and 0.85$R_{\odot}$.

\subsection{Summary}\label{Sec:Summary}

\begin{figure}
	\centering
		\includegraphics[trim= 0 0 0 0, clip, width=9cm]{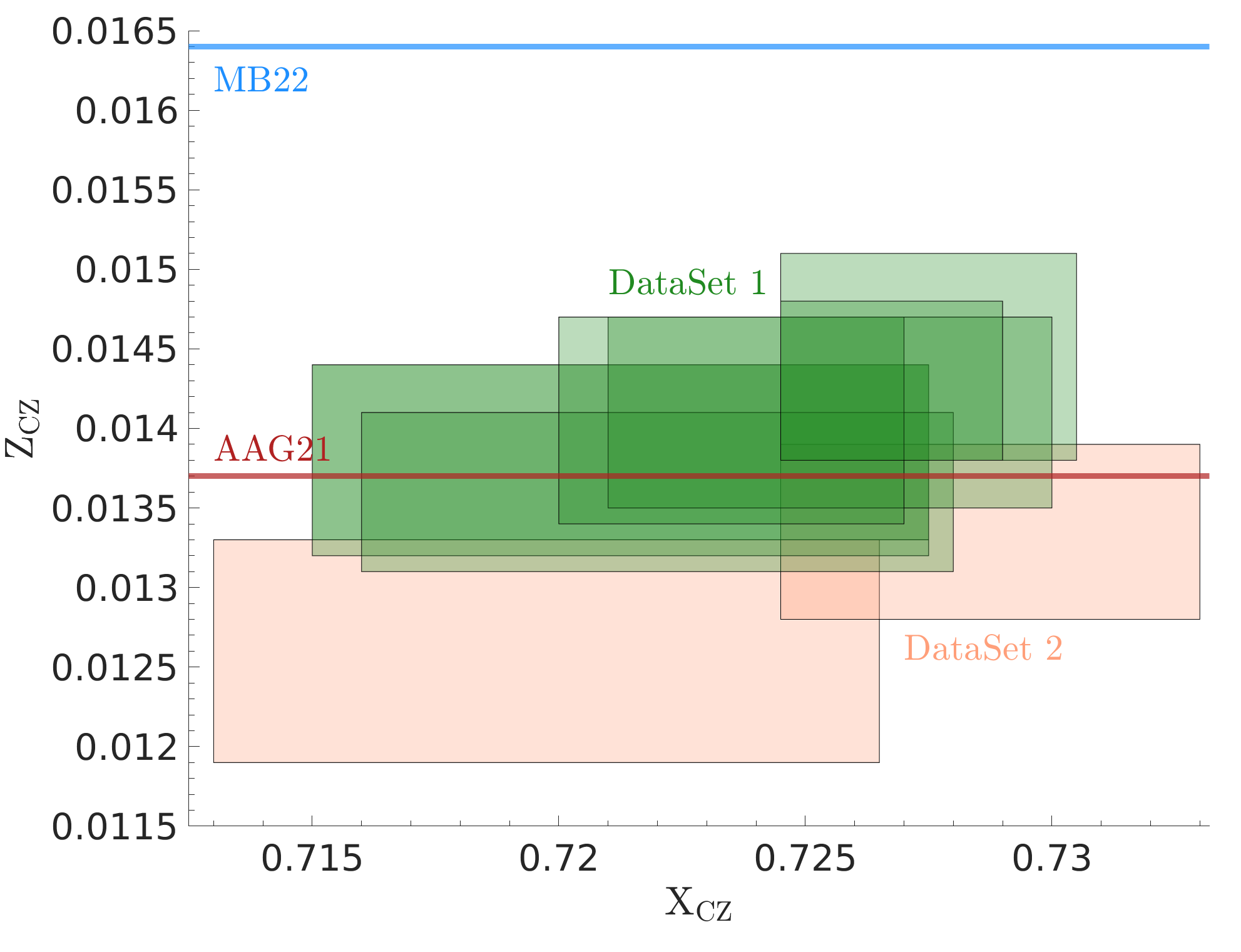}
	\caption{Summary of the inversion procedure for the various models, datasets and equations of state. The green and red rectangles represent the intervals provided from the analysis of Sect. \ref{Sec:RealData} and \ref{Sec:Data}, respectively. The blue and red lines indicate the solar metallicity value from \citet{Magg2022} (MB22) and \citet{Asplund2021} (AAG21), respectively.}
		\label{Fig:Summary}
\end{figure}

To provide a global view of the inversion results, we have to combine the information of the $\chi^{2}$ for both datasets and all test cases combined. This is done in Fig. \ref{Fig:Summary} and in Table \ref{tabZSummary}. In this table, each line represents a full inversion procedure, from the determination of the seismic model, to the reconstruction of the $\Gamma_{1}$ profile and chemical composition determinations. Some cases, in line 5 and 6, are those for which a different equation of state than that of the reference model was used in the $\Gamma_{1}$ reconstruction procedure. We chose to use the limits of the white areas in the $\chi^{2}$ maps of the hydrogen determinations to derive confidence intervals to be used to extract the associated metallicity intervals. As shown in Sect. \ref{Sec:HH}, this allowed to recover the accurately actual values in the exercices with artifical data. 

\begin{table*}[h]
\caption{Summary of the inversion results for various equations of state and dataset}
\label{tabZSummary}
  \centering
\begin{tabular}{r | c | c | c | c }
\hline \hline
\textbf{Model}&\textbf{EOS}&\textbf{Dataset}&\textbf{X interval}&\textbf{Z interval} \\ \hline
Model A1 & FreeEOS & 1  & $0.728\pm 0.003$ & $0.0145\pm 0.0007$\\
Model A2 & SAHA-S v7 & 1 & $0.721\pm 0.006$ & $0.0139\pm 0.0006$ \\ 
Model A3 & SAHA-S v3 & 1 & $0.721\pm0.006$ & $0.0138\pm 0.0006$\\
Model M1 & FreeEOS & 1 & $0.727\pm0.003$ & $0.0143\pm0.0005$ \\ 
Model M1 (SAHA-S v7) & SAHA-S v7 & 1 & $0.724\pm0.004$ & $0.0141\pm 0.0006$\\
Model A2 (FreeEOS) & FreeEOS & 1 & $0.726\pm0.005$ & $0.0141\pm 0.0006$\\
Model M2 & SAHA-S v7 & 1 & $0.722\pm0.006$ & $0.0136\pm 0.0005$\\
Model M1 & FreeEOS & 2 & $0.720\pm0.007$& $0.0126\pm0.0007$ \\
Model A2 & SAHA-S v7 & 2 & $0.729\pm0.005$ & $0.0134\pm 0.0006$\\
\hline
\end{tabular}
\end{table*}

It is to be noted that the size of the rectangles will depend on other parameters such as data set and equation of state, as this will change the landscape of the $\chi^{2}$. Our approach shows that the metallicity interval determined from a detailed helioseismic analysis of the properties of the solar envelope strongly favours the \citet{Asplund2021} abundances over the \citet{Magg2022} abundances.

\section{Conclusion}\label{Sec:Conc}

The conclusions of this study is straightforward, inversions of solar data to determine the $\Gamma_{1}$ profile in the solar convective envelope do not favour the revision of the abundances of \citet{Magg2022} back to the old \citet{GS1998} metallicity value. As seen from Fig. \ref{Fig:Summary} this independent measurement of the solar metallicity from seismic analyses of the solar envelope strongly favours AAG21, as well as a high Y value in the convective envelope. The situation for the \citet{Magg2022} abundances ais actually worse because the strongest rejection is found for a high Z, high Y model, which corresponds to the output of a MB22 model reproducing the solar luminosity. The situation worsens further for MB22 models including macroscopic transport to reproduce the lithium depletion at the age of the Sun \citep{Buldgen2023}. 

Regarding helium, it also appears that the high Y value favoured here cannot be attained solely by including the effects of macroscopic transport in low-Z models. A revision of nuclear reaction rates or opacities at high T is required to reconcile solar models with the analysis performed here \citep{Ayukov2017}.

The method implemented here shows a few caveats. The inversion is quite difficult, as trade-off parameters have to be adjusted for each dataset and artificial data plays a key role in the calibration. Therefore the results cannot be obtained on a large scale with numerous datasets. From supplementary investigations, suboptimal trade-off parameters do not change the conclusions, but rather push towards lower Z values. Testing on more datasets might be done incrementally in the future but seems unlikely to change the conclusions. Future revisions of the EOS, or availability of new tables would be interesting to test physical effects in the EOS and further confirm our results\footnote{For example the latest revision of the MHD equation of state (R.Trampedach private communication)}. An additional limitation of the method is the treatment of surface effects, here chosen to be dealt with using a $6^{th}$ order polynomial form as in \citet{RabelloParam}. Experiments with both artificial data and changes to the order of the polynomial correction have been conducted to ensure robustness, but pushing towards higher degrees might need more detailed functional forms \citep{DiMauro2002} and the robustness and precision of the method would be further improved by using improved modelling approach of the surface layers \citep{Spada2018,Jorgensen2021}. Further investigations on the systematics of such inversions are therefore required to further pin down the chemical composition of the solar envelope, as well as detailed comparisons of the available equations of state of the solar material.

Nevertheless, from our detailed helioseismic analysis of the solar envelope using an advanced seismic inversion approach and up-to-date equations of state of the latest generations of solar models, we conclude that the solar metallicity in the convective envelope lies in the range $0.0120-0.0151$ and the solar hydrogen mass fraction in the envelope lies in $0.715-0.730$, resulting in a $(Z/X)_{\odot}$ value between $0.0168-0.0205$. Moreover, high metallicity models using the \citet{Magg2022}, \citet{GS1998} or \citet{GrevNoels} abundances are rejected as a steep slope in $\chi^{2}$ values is observed in all cases, with the $\chi^{2}$ values of high-metallicity solar envelope models being a factor 6 to 20 higher than those of low-metallicity models, independently of the EOS used and for two different helioseismic datasets. This independent measurement of the solar metallicity therefore strongly supports the AAG21 abundances \citep{Asplund2021} over the MB22 abundances \citep{Magg2022}, in line with previous studies using modern equations of state \citep{Vorontsov13,VorontsovSolarEnv2014,BuldgenZ}. Compared to these previous studies, our method provides a much more robust and precise inference, exploiting the properties of existing equations of state of the solar material. 

\section*{Acknowledgements}

The authors thank the referee for their careful reading of the manuscript. G.B. is funded by the SNF AMBIZIONE grant No 185805 (Seismic inversions and modelling of transport processes in stars). A.M.A. gratefully acknowledges support from the Swedish Research Council (VR 2020-03940). We acknowledge support by the ISSI team ``Probing the core of the Sun and the stars'' (ID 423) led by Thierry Appourchaux. The authors thank J. Christensen-Dalsgaard and S. Vorontsov for the detailed reading of the manuscript and the numerous suggestions. 

\bibliography{biblioarticleZArt}

\newpage
\begin{appendix}
\section{Additional tests}

In addition to the tests presented in Sect. \ref{Sec:RealData}, we also performed additional inversions. First, we used model M1, computed with FreeEOS, combined with the SAHA-S v7 equation of state in the last reconstruction step. Second, we used model A2, computed with SAHA-S v7, combined with FreeEOS in the reconstruction. 

\begin{figure*}
	\centering
		\includegraphics[trim= 0 0 0 0, clip, width=16cm]{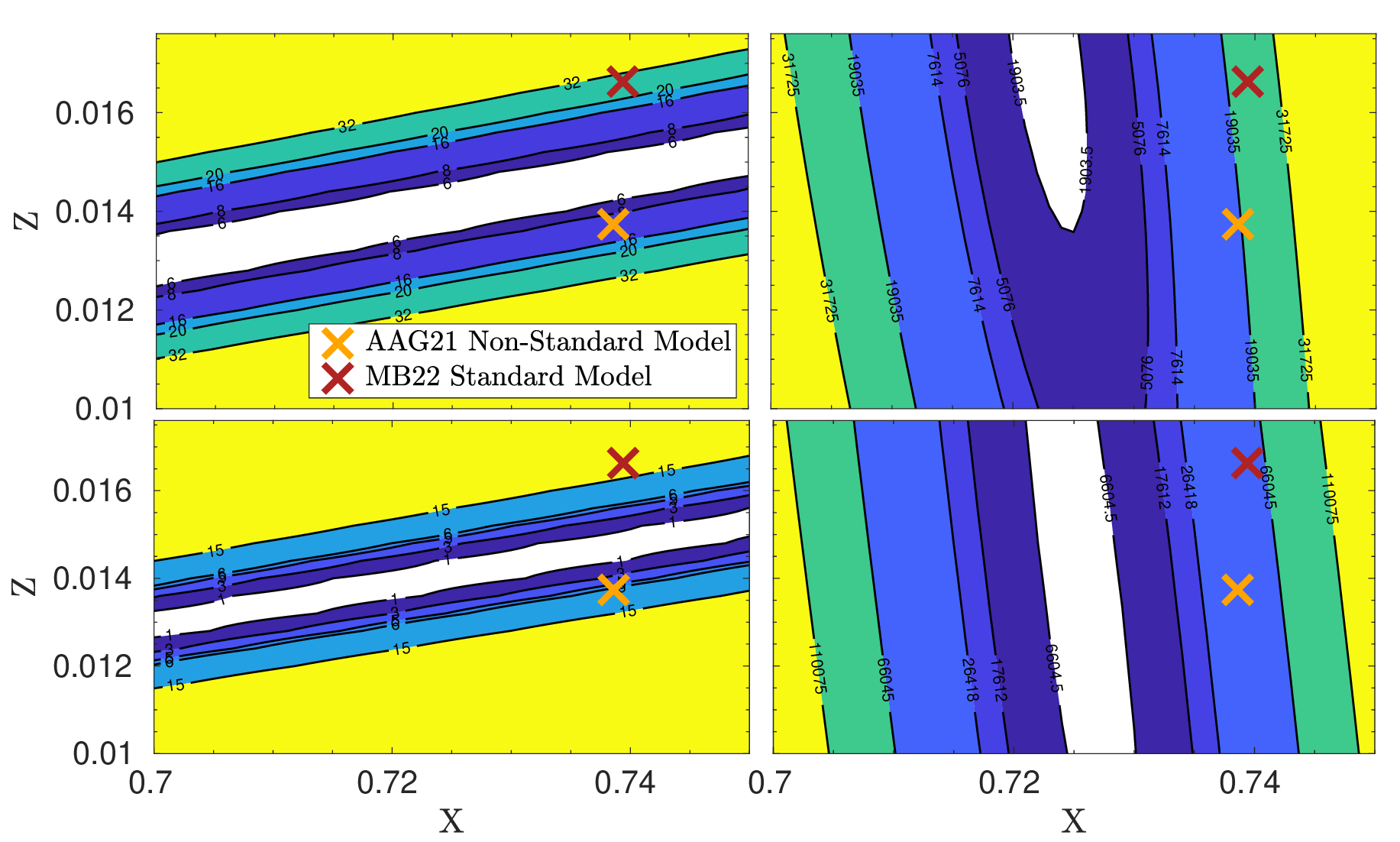}
	\caption{$\chi^{2}$ map of the X and Z scan for solar data. The orange and red crosses indicate the positions of AAG21 model including rotation and magnetic fields and the MB22 standard solar model respectively (values from their paper). The upper panels are associated with model M1 and the lower panels are associated with model A2 (see text for details). The lefet panels are associated with the high-T subdomain for the Z determination, while the right panels is associated with the low-T subdomain used to determine X.}
		\label{Fig:InvSunM1A2}
\end{figure*}

The results are illustrated in Fig. \ref{Fig:InvSunM1A2}, with the lower panels being associated with the reconstruction starting from A2 and the lower panels are associated with the results using M1. The results are unchanged with respect to the solutions found previously, meaning that the final reconstructed $\Gamma_{1}$ is not affected by the initial EOS used in the reference model. It does however depend on the EOS used in the last reconstruction step that determines the values of X and Z in the solar envelope.

\section{Solar data}

Table \ref{tabModeSet} summarizes the sets of modes used in this paper. The full dataset is provided at \textbf{Add URL for online material}. BiSON data only covers low $\ell$ modes from 0 to 3, while MDI data is used for all higher degrees.

\begin{table*}[h]
\caption{Datasets used in the inversion procedure}
\label{tabModeSet}
  \centering
\begin{tabular}{r | c | c | c  }
\hline \hline
\textbf{Instruments}&\textbf{References}&\textbf{Degrees}\\ \hline
BiSON + MDI & \citet{Davies},\citet{BasuSun} & 0 - 250\\
BiSON + MDI & \citet{Davies},\citet{BasuSun},\citet{Larson2015} & 0 - 250 \\ 
\hline
\end{tabular}
\end{table*}

\end{appendix}

\end{document}